\documentclass[twocolumn,10pt]{IEEEtran}
\usepackage{amsmath,amssymb}
\usepackage{graphicx}
\usepackage{epstopdf}
\usepackage{pgf}
\usepackage{tikz}
\usepackage{cite}
\usepackage{setspace}
\usepackage{pgffor}
\usepackage{pgfplots}
\usepackage{pgfplotstable}
\usetikzlibrary{arrows,shapes,positioning,plotmarks,fit}
\usepgflibrary{shapes.geometric}
\newcommand{\norm}[1]{\left\|{#1}\right\|}

\def\zbf{\mathbf{z}}

\def\zbf{\mathbf{z}}

\newtheorem{theorem}{Theorem}

\newtheorem{lemma}{Lemma}

\makeatletter
\def\mod#1{\allowbreak\mkern6mu{\operator@font mod}\,\,#1}
\makeatother

\title{Toward Resource-Optimal Consensus over the Wireless Medium}

\author{
\IEEEauthorblockN{Matthew~Nokleby, Waheed U. Bajwa, Robert Calderbank, and Behnaam Aazhang} \thanks{Copyright (c) 2013 IEEE. Personal use of this material is permitted. However, permission to use this material for any other purposes must be obtained from the IEEE by sending a request to pubs-permissions@ieee.org

M. Nokleby and R. Calderbank are with Duke University, Durham, NC (email: \{matthew.nokleby,robert.calderbank\}@duke.edu). W. U. Bajwa is with Rutgers University, Piscataway, NJ (email: waheed.bajwa@rutgers.edu). B. Aazhang is with Rice University, Houston, TX (email: aaz@rice.edu). This work was supported by the National Science Foundation, the Academy of Finland, the Tekes FiDiPro program, and the Air Force Office of Scientific Research under grant  FA 9550-09-1-0643.
Portions of this work were presented at the International Conference on Acoustics, Speech, and Signal Processing, Mar. 2012 and the Asilomar Conference on Signals, Systems, and Computers, Nov. 2012.}
}

\begin{document}
\maketitle
%\doublespacing
%\onehalfspacing

\begin{abstract}
We carry out a comprehensive study of the resource cost of averaging consensus in wireless networks. Most previous approaches suppose a graphical network, which abstracts away crucial features of the wireless medium, and measure resource consumption only in terms of the total number of transmissions required to achieve consensus. Under a path-loss dominated model, we study the resource requirements of consensus with respect to three wireless-appropriate metrics: total transmit energy, elapsed time, and time-bandwidth product. First we characterize the performance of several popular gossip algorithms, showing that they may be order-optimal with respect to transmit energy but are strictly suboptimal with respect to elapsed time and time-bandwidth product. Further, we propose a new consensus scheme, termed {\em hierarchical averaging}, and show that it is nearly order-optimal with respect to all three metrics. Finally, we examine the effects of quantization, showing that hierarchical averaging provides a nearly order-optimal tradeoff between resource consumption and quantization error.
\end{abstract}
%\allowdisplaybreaks[3]

\section{Introduction}
 Consider a network of $N$ nodes, each of which possesses a scalar measurement $z_n(0) \in \mathbb{R}$. In  {\em averaging consensus}, each node wishes to compute the average of these measurements: 
$ 	z_{\mathrm{ave}} = 1/N \sum_{n=1}^N z_n(0).$ Averaging consensus often is described as a simple, canonical example of distributed signal processing over sensor networks; a common narrative is that each node measures the local temperature and wants to compute the average temperature over the sensor field. Such simplicity, however, is deceptive, as consensus lies at the heart of an array of sophisticated problems, including load-balancing \cite{cybenko:JPDC89}, distributed optimization \cite{tsitsiklis:PHD84,tsitsiklis:TAC86,duchi:TAC12}, and distributed estimation and filtering \cite{saligrama:TSP06,spanos:ISIPSN}. Advances in consensus therefore represent advances throughout distributed signal processing. 
 
Due to their simplicity, flexibility, and robustness, {\em gossip algorithms} have emerged as a popular approach to consensus. In gossip, the network is modeled by a graph. Nodes iteratively pair with neighbors, exchange estimates, and average those estimates together, eventually converging on the true average. A large body of excellent work on gossip has been developed, from the early {\em randomized gossip} of \cite{boyd:IT06} to faster schemes such as {\em path averaging} \cite{benezit:IT10} and {\em multi-scale gossip} \cite{tsianos:IT10}. Gossip is simple, requiring minimal processing and network knowledge, and it is robust, retaining performance even with failing links and dynamic topology.

%nodes iteratively pair with neighbors, exchange estimates, and average those estimates together, eventually converging on the true average with high probability. Gossip is simple, requiring minimal processing and network knowledge, and it is robust, retaining performance even when communications links fail or network topology changes. A large body of excellent work on gossip has been developed, from the {\em randomized gossip} \cite{boyd:IT06} of Boyd et al., which initiated recent interest in gossip, to {\em path averaging} gossip \cite{benezit:IT10} of Benezit et al., which is order-optimal in convergence speed, to {\em quantized consensus} \cite{kashyap:auto07} of Kashyap et al., which addresses consensus over networks with finite-capacity links.

%Many variations on gossip have been proposed in the literature, from the {\em randomized gossip} \cite{boyd:IT06}, which inaugurated recent interest in gossip, to more sophisticated approaches, such as {\em path-averaging}, which are order-optimal in convergence speed, requiring only $O(N)$ transmissions\footnote{Throughout this paper we use the Landau notation: $f(n) = O(g(n))$ implies $f(n) \leq kg(n)$, $f(n) = \Omega(g(n))$ implies $f(n) \geq k g(n)$, and $f(n)= \Theta(g(n))$ implies $f(n) = O(g(n))$ and $f(n) = \Omega(g(n))$, all for constant $k$ and sufficiently large $n$.} to achieve consensus over random networks.

However, the purpose of consensus strategies is often to facilitate processing over {\em wireless} networks, and wireless affords possibilities that existing strategies do not fully exploit. For example, in random geographic graphs, nodes are taken to be neighbors if they lie within a certain radius, usually chosen to be $\Theta(\sqrt{\log{N}/N})$ following \cite{gupta:CDC98}. Sophisticated consensus algorithms are then constructed in order to minimize the number of transmissions needed to achieve consensus. In wireless, however, transmit radius is adjustable via power control. Furthermore, given sufficient transmit power the network is fully connected, at which point consensus is trivial. This suggests both that wireless permits flexibility that may improve performance and that we must consider additional performance metrics---such as transmit power---that encompass more than just the number of transmissions.

%Based on the preceding, we make several observations which motivate this work:
%\begin{itemize}
%	\item Consensus algorithms are typically designed over graphical networks, which abstract away the broadcast and superposition nature of the wireless medium. In wireless, a single transmission arrives at multiple destinations, and multiple transmissions interfere at a single destination. By contrast, transmissions over graphical networks are usually assumed to be unilateral.
%	\item Graphical networks presuppose a fixed topology, whereas in wireless networks connectivity adjusted dynamically via the transmit power.
%	\item The performance of consensus algorithms typically is measured with respect to the total number of transmissions. In wireless, however, multiple resources are expended---namely time, bandwidth, and energy---which must be taken into consideration.
%\end{itemize}

While these issues have been taken up individually, in this work we consider them jointly to answer the following question: {\em What are the resource demands of consensus over wireless  networks?} Our objective is to expand the framework in which consensus is studied in order to account for and exploit features of the wireless medium that previously have been overlooked. Given the ability to broadcast  and to adjust connectivity dynamically, we seek fundamental limits on the wireless resources required to achieve consensus, as well as practical consensus strategies that attain those limits.

\subsection{Contributions}
First, in Section \ref{sect:system.model} we define a realistic but tractable framework in which to study the resource demands of consensus. We assume a path-loss dominated propagation model in which connectivity is determined by a signal-to-noise ratio (SNR) threshold. 
%At first we suppose ideal wireless links for which quantization noise is negligible.
We define three resource metrics: the total energy expended in order to achieve consensus, the total time elapsed, and the time-bandwidth product consumed. In Section \ref{sect:inner.bounds} we derive lower bounds on the required resources, and in Section \ref{sect:gossip.performance} we characterize several existing consensus strategies within our framework. We show that while path averaging is nearly order optimal with respect to energy expenditure, it remains suboptimal with respect to elapsed time and consumed time-bandwidth product.

Next, in Section \ref{sect:hierarchical.averaging} we propose a new %\footnote{An early version of hierarchical averaging was presented in \cite{nokleby:ICASSP12}.} 
consensus algorithm, termed {\em hierarchical averaging}, designed specifically for wireless networks. Instead of communicating with neighbors over a graph, nodes broadcast estimates to geographically-defined clusters. These clusters expand as consensus proceeds, which is enabled by adjusting nodes' transmit powers. Similar to the hierarchical cooperation of \cite{ozgur:IT07} and the multiscale gossip of \cite{tsianos:IT10}, small clusters cooperatively broadcast information to larger clusters, continuing until consensus is achieved. Depending on the particulars of the channel model, hierarchical averaging is nearly order optimal. In particular, when channel phases are fixed and identical, hierarchical averaging is order optimal with respect to all three metrics simultaneously, up to an arbitrarily small gap in the exponent, for path-loss exponents $2 \leq \alpha < 4$. In the more realistic case in which phases are random and independent, however, hierarchical averaging is no longer order optimal in transmit energy when $\alpha > 2$, although it remains order optimal with respect to the other two metrics.

Finally, in Section \ref{sect:quantization} we incorporate quantization into our model. Since practical wireless links suffer from noise, achievable rates are finite and estimates must be quantized prior to transmission. This introduces a tradeoff: expending more energy increases the rate of the links, thereby reducing the quantization error inherent to each transmission and therefore the estimation error accrued during consensus. Therefore, in addition to the resource metrics of energy, time, and bandwidth, we introduce a fourth performance metric: mean-square error of nodes' estimates. Again we characterize existing consensus techniques. We also apply quantization to hierarchical averaging, showing that it permits an efficient tradeoff between energy and estimation error.
\subsection{Prior Work}
% detail a few works that are relevant to the present study. For a comprehensive examination of consensus and gossip, see the excellent survey \cite{dimakis:IEEE10} of Dimakis et al.

Consensus has been studied under various guises, including the early work of Tsitsiklis \cite{tsitsiklis:PHD84}, which examined averaging in the context of distributed estimation. Recent interest in consensus was sparked by the introduction of randomized gossip \cite{boyd:IT06}, which defined the framework and developed the theoretical machinery in which most subsequent works have operated. %A similar deterministic consensus was proposed in \cite{xiao:SCL04}.
Randomized gossip, however, has relatively slow convergence on random graphs, requiring roughly $\Theta(N^2)$ transmissions.\footnote{Throughout this paper we use the Landau notation: $f(n) = O(g(n))$ implies $f(n) \leq kg(n)$, $f(n) = \Omega(g(n))$ implies $f(n) \geq k g(n)$, and $f(n)= \Theta(g(n))$ implies $f(n) = O(g(n))$ and $f(n) = \Omega(g(n))$, all for constant $k$ and sufficiently large $n$.} Since then, researchers have searched for faster consensus algorithms. In {\em geographic gossip} \cite{dimakis:TSP08}, nodes pair up with geographically distant nodes, exchanging estimates via multi-hop routing. The extra complexity garners faster convergence; geographic gossip requires roughly $\Theta(N^{3/2})$ transmissions. Geographic gossip was further refined by the introduction of {\em path averaging} \cite{benezit:IT10}, in which routing nodes contribute their own estimates ``along the way.'' Path averaging closes the gap to order optimality, requiring roughly $\Theta(N)$ transmissions, which is the minimum of any consensus algorithm.

A few works have addressed individually the wireless aspects we consider herein. The broadcast nature of wireless is considered in \cite{ustebay:Allerton08,aysal:TSP09}; however, in these works broadcast does not significantly improve performance over randomized gossip. % In fact, in \cite{aysal:TSP09} the required number of transmissions scales {\em worse} than in randomized gossip.
Multi-access interference is addressed---and in fact exploited---in \cite{nazer:STSP11}, where lattice codes are used to compute sums of estimates ``over the air.'' %This obviates the need for interference mitigation, but it applies only to very specific network topologies.
In works that presage our own \cite{sardellitti:SP12,vanka:JSCC10,vanka:JSTSP11}, the effects of power control are explored. In \cite{sardellitti:SP12} the energy-minimal {\em fixed} power allocation for traditional consensus is derived. In \cite{vanka:JSCC10,vanka:JSTSP11} the convergence speed of traditional consensus is studied as a function of nodes' fixed transmit radius, supposing that TDMA is used to mitigate interference among users.

Finally, many authors have studied the impact of noisy links on consensus. In \cite{xiao:JPDC07}, continuous-valued estimates are corrupted by zero-mean additive noise, and optimal linear consensus strategies are derived. For a similar model, the {\em bias-variance dilemma} is identified: Running consensus for longer reduces the bias of the resulting estimates, but it increases the variance. %Algorithms that resolve the dilemma are presented, but they suffer from slow convergence.
In \cite{kashyap:auto07,benezit:JSAC11} {\em quantized consensus} algorithms are presented that achieve consensus over finite alphabets. In \cite{aysal:SP08} standard gossip is augmented with {\em dithered} quantization and are shown to achieve consensus on the true average in expectation. In \cite{yildiz:SP08} the increasing correlation among estimates is exploited to construct a consensus algorithm employing Wyner-Ziv style coding with side information.

%\subsection{Organization}
%In Section \ref{sect:quantization.metrics} we detail the wireless model to be used throughout this work, and we formalize the performance metrics by which we will characterize consensus strategies. In Section \ref{sect:inner.bounds} we derive lower bounds on the resource requirements of consensus under our model. In Section \ref{sect:gossip.performance} we characterize the resource requirements existing consensus strategies, assuming that links are perfect and have infinite rate. In Section \ref{sect:hierarchical.averaging} we introduce hierarchical averaging and characterize its resource requirements assuming infinite-rate links. Then, in Section \ref{sect:quantization} we consider quantization, characterizing the resource/estimate tradeoff of existing approaches as well as hierarchical averaging. Finally, we conclude with Section \ref{sect:conclusion}.

\section{Preliminaries}\label{sect:system.model}
\subsection{System Model}\label{sect:sub.system.model}
In defining the wireless model, we aim for a balance between tractability and practicality. To this end we make four critical assumptions, which we contend capture the salient features of wireless while maintaining simplicity: synchronous transmission, path-loss propagation, ``protocol''-model connectivity, and orthogonalized interference management. %In this subsection we detail and justify our assumptions.

%Although consensus algorithms are occasionally defined over synchronous models (e.g., the synchronized gossip from \cite{boyd:IT06}), researchers more commonly assume communications to be asynchronous. Each node has an independent clock that ``ticks'' at Poisson-distributed intervals; upon each clock tick the node initiates a round of consensus, which is assumed to take place instantaneously. This model is an idealized version of ALOHA-style protocols, and it sidesteps the scheduling and interference difficulties inherent to wireless communications. Our goal, however, is both to characterize the best possible performance under wireless and to address interference. We therefore adopt a synchronous model in which nodes transmit simultaneously in slotted time. In practice, near-perfect synchronization can be achieved via beacons, as in superframes for 802.15, or via GPS clocks.

\subsubsection{Synchronous Transmission}
Let transmissions be broken up into time slots $t$, each slot having $K$ channel uses. At time slot $t$, node $n$ transmits a signal having average power $P_n(t)$. Nodes use capacity-achieving codes, and $K$ is chosen sufficiently long that the probability of error is negligible.

\subsubsection{Path-loss Propagation}
Each node $n$ has a geographic location $\mathbf{r}_n \in [0,1] \times [0,1]$ in the unit square, which we take to be independently drawn from a uniform distribution. Let $1\leq k \leq K$ be a symbol time during block $k$. Under the path-loss model, the channel gain between any two nodes $m,n$ is
\begin{equation}\label{eqn:channel.gains}
	h_{mn}(t,k) = \sqrt{G}\norm{\mathbf{r}_m - \mathbf{r}_n}_2^{-\alpha/2}e^{j\theta_{mn}(t,k)},
\end{equation}
where $\alpha \geq 2$ is the path-loss exponent, $\theta_{mn}(t,k) \in [0, 2\pi)$ is a random phase with unspecified distribution, and $G$ is a constant that depends on $\alpha$ and the transmit frequency. The choice of $G$ does not affect scaling laws. In simulations, we choose $G$ sufficiently small that channel gains are smaller than unity.

\subsubsection{Protocol Model}
Following the channel model in (\ref{eqn:channel.gains}), the signal transmitted by node $m$ arrives at node $n$ with average power
\begin{equation}
	R_{mn}(t) = \frac{1}{K}\sum_{k=1}^K |h_{mn}(t,k)|^2P_n(t).
\end{equation}
%Since the network resides on the unit square, in practice $G$ is smaller than unity to ensure that the channel gains attenuate signals rather than amplify them. Furthermore, in practice $G$ varies with the path-loss exponent to ensure that channel magnitudes decrease with increasing path-loss. For concreteness, and since we are concerned primarily with scaling laws, however, we choose $G=1$. This choice does not affect the scaling laws, but it may give the false impression that resource requirements decrease with increasing path-loss exponents. While we shall see that scaling laws in $N$ appear more favorable for higher $\alpha$, in practice the absolute resource requirements for any fixed $N$ increase with $\alpha$.
We suppose that a link exists between node $n$ and $m$ provided the received power is above a signal-to-noise ratio threshold $\gamma > 0$.  Define the {\em neighborhood} of node $n$ as the set of nodes whose transmissions have sufficient received power:
\begin{align}
	\mathcal{N}_n(t) &= \{m : R_{mn}(t) \geq \gamma \} \notag \\
	&= \left\{m : P_m(t) \geq \frac{\gamma}{G} \norm{\mathbf{r}_m - \mathbf{r}_n}_2^\alpha \right\}. \label{eqn:power.radius}
\end{align}
For nodes $m \notin \mathcal{N}_n(t)$, we assume that node $n$ suffers no interference from node $m$'s transmission. This assumption permits a tractable, geometric analysis of connectivity.

In hierarchical averaging, presented in Section \ref{sect:hierarchical.averaging}, we group nodes into clusters which transmit cooperatively. In this case we must expand the definition of neighborhoods to characterize the number of unique {\em signals} arriving at node $n$. Let $\mathcal{C}_{i} \subset \{1,\cdots,N\}$, where $i \in \mathcal{I}$ for some index set $\mathcal{I}$, form a partition of the network. At time slot $t$, nodes within cluster $\mathcal{C}_j$ transmit jointly. Define the received power at node $n$ as
\begin{align*}
	R_{\mathcal{C},n}(t) &= \frac{1}{K}\sum_{k=1}^K\left|\sum_{m \in \mathcal{C}_j}h_{mn}(t,k)\sqrt{P_m(t)}\right|^2 \\
	&\approx E\left[\left|\sum_{m \in \mathcal{C}_j}h_{mn}(t,k)\sqrt{P_m(t)}\right|^2\right],
\end{align*}
where the expectation is taken over the random phases, and the approximation is accurate for large $K$. Then, the neighborhood of $n$ is the set of all clusters $\mathcal{C}_j$ such that the received power exceeds $\gamma$:
\begin{align*}
	\mathcal{N}_{n}(t) %&=  \left\{ \mathcal{C} : R_{\mathcal{C},n}(t) \geq \gamma \right\} \\
	&= \left\{ \mathcal{C}_j : E\left[\left|\sum_{m \in \mathcal{C}_j}h_{mn}(t,k)\sqrt{P_m(t)}\right|^2\right] \geq \gamma,  \right\}.
\end{align*}
The connectivity of clusters depends on the distribution of the phases $\theta_{mn}(t,k)$. In the sequel we consider two possibilities. First, we consider the simple case in which the phases are equal and fixed. In this case, signals constructively combine at receivers, and the neighborhood of $n$ can be written as
 \begin{equation}\label{eqn:coherent.neighborhoods}
	\mathcal{N}_n(t) = \left\{\mathcal{C}_j : \left(\sum_{m \in \mathcal{C}_j} \norm{\mathbf{r}_m - \mathbf{r}_n}_2^{-\alpha/2} P^{\frac{1}{2}}_m(t)\right)^2 \geq \frac{\gamma}{G} \right\}.
\end{equation}
The second, and more realistic, case we consider is that each $\theta_{mn}(t,k)$ is independently and uniformly distributed across $[0,2\pi)$. Then, signals do not combine coherently, and the neighborhood of $n$ is
 \begin{equation}\label{eqn:non.coherent.neighborhoods}
	\mathcal{N}_n(t) = \left\{\mathcal{C}_j :\sum_{m \in \mathcal{C}_j} \norm{\mathbf{r}_m - \mathbf{r}_n}_2^{-\alpha} P_m(t) \geq \frac{\gamma}{G} \right\}.
\end{equation}
Finally, we define the neighborhood of a {\em cluster} $\mathcal{C}_i$ as those clusters in the neighborhood of any of its nodes:
\begin{equation}
	\mathcal{N}_{\mathcal{C}_i}(t) = \{ \mathcal{C}_j : \exists n \in \mathcal{C}_i \text{ with } \mathcal{C}_j \in \mathcal{N}_n(t) \}
\end{equation}

\subsubsection{Frequency Orthogonalization}
%Our final assumption is a simple orthogonalized approach to interference management.
At every time slot $t$, each node $n$ transmits over a frequency slot $f_n(t) \in \mathbb{N}$. In order for transmissions to be successful, each incoming signal must arrive on a different frequency slot, which means that each node must transmit on a  frequency different not only from its neighbors, but also from its neighbors' neighbors. The number of required frequency slots can be described in terms of graph coloring. Let $\mathcal{G}_2(t)$ denote the graph of two-hop neighbors---of either nodes or clusters, as appropriate---at time slot $t$. Then, the number of frequency slots required at time $t$ is
\begin{align*}
	B(t) = \chi(\mathcal{G}_2(t)),
\end{align*}
where $\chi(\cdot)$ is the (vertex) chromatic number. Clearly,
\begin{equation}\label{eqn:bandwidth.min}
	B(t) \geq \max_n |\mathcal{N}_n(t)|
\end{equation}
Further, by Brooks' theorem \cite{brooks:CPS41},
\begin{equation}\label{eqn:bandwidth.max}
	B(t) \leq \max_n \mathrm{Deg}_n(\mathcal{G}_2(t))+1,
\end{equation}
where $\mathrm{Deg}_n(\cdot)$ is the {\em degree} of node $n$ over $\mathcal{G}_2(t)$, or the number of two-hop neighbors.

\subsection{Performance Metrics: Ideal Links}
We first consider the case in which the links between neighboring nodes are ideal, meaning that the quantization artifacts associated with finite-rate transmission are neglected. Therefore, at time $t$, node $n$ decodes an infinite-precision, real-valued scalar from each $m \in \mathcal{N}_n(t)$. %This simplification is a common one in consensus algorithms, even those that address aspects of the wireless medium \cite{vanka:JSTSP11}.
%Later we will examine the effects of non-ideal links requiring quantization.

At time $t$, each node updates its estimates by taking a linear combination of estimates received from neighboring nodes:
\begin{equation}
	z_n(t) = \sum_{m \in \mathcal{N}_n(t)} a_{mn}(t) z_m(t),
\end{equation}
where $a_{mn}(t)$ are arbitrary  coefficients. Note that the  coefficients, as well as the neighborhoods, may vary with time. Since connectivity may vary in time, the weights do not necessarily need to form a doubly-stochastic matrix, as is typically required in consensus with fixed topologies \cite{boyd:IT06}.
%Of course, a poor choice of coefficients may result in slow convergence or no convergence at all.

The first performance metric is the {\em $\epsilon$-averaging time}, defined as the number of time slots required to achieve consensus to within a specified tolerance\footnote{This definition is inspired by the well-known connection between Markov chain mixing and averaging consensus \cite{xiao:SCL04}. }:
\begin{equation}\label{eqn:averaging.time}
	T_\epsilon = \sup_{\zbf(0) \in \mathbb{R}^n}\inf\left\{ t : \mathrm{Pr}\left(\frac{\norm{\zbf(t)-z_{\mathrm{ave}}\mathbf{1}}}{\norm{\mathbf{z}(0)}} \geq \epsilon \right) \leq \epsilon \right\},
\end{equation}
where $\mathbf{z}(t)$ is the vector of estimates $z_n(t)$. The scaling law of $T_\epsilon$ is the primary focus of study in consensus. However, it provides only a partial measure of resource consumption in wireless networks, so we consider other metrics as well.
%For example, as mentioned in the introduction, with enough energy and bandwidth it is trivial to achieve perfect consensus in a single time slot. We therefore study these resources as well.

%We next examine energy, which is scarce in networks composed of cheap, battery-powered nodes. Define
The next metric is
the {\em total transmit energy}, the energy required to achieve consensus to within the tolerance $\epsilon$:
\begin{equation}
	E_\epsilon = K \sum_{n=1}^N\sum_{t=1}^{T_\epsilon} P_n(t).
\end{equation}
%Supposing each channel use to have unit time and each frequency slot to have unit bandwidth, $E_\epsilon$ gives the energy, in Joules, expended by the network.
%Supposing each time slot to be of equal length, the transmit power $P_n(t)$ is proportional to the energy consumed by node $n$ over slot $t$. Summing over nodes and time slots yields the total energy consumed.

The final figure of merit is the {\em time-bandwidth product}, the total number of frequency slot uses required to achieve consensus to within the tolerance $\epsilon$:
\begin{equation}\label{eqn:time.bandwidth}
	B_\epsilon = K\sum_{t=1}^{T_\epsilon} B(t),
\end{equation}
%The metric $B_\epsilon$ gives the total number of frequency slot uses required to achieve consensus to within tolerance $\epsilon$.
%As mentioned previously, we leave unspecified whether the sub-channels are realized in time, frequency, or code. However $T_\epsilon$ represents the {\em temporal component} of the time-bandwidth product. The sequential nature of consensus dictates that $T_\epsilon$ rounds occur in succession. Therefore $T_\epsilon$ characterizes a constraint on the realization of the time-bandwidth product. All of the time-bandwidth product may be realized with temporal resources, but only a fraction of it may be realized by frequency resources.

%The definition of these three metrics suggests a trade-off in which improvement with respect to one metric entails degraded performance with respect to the others. In the sequel, however, we will show that simultaneously near-optimal performance is possible with respect to all three metrics under certain circumstances.

\subsection{Performance Metrics: Finite-Rate Links}\label{sect:quantization.metrics}
In practice, wireless links are noisy and therefore have finite rate, which precludes the infinite-precision exchange of scalars. Instead, nodes must quantize their estimates to a finite alphabet prior to each round of consensus. To simplify the discussion, we suppose that the measurements $z_n(0)$ are drawn from the finite interval $[0,1)$. Throughout this paper, we employ dithered uniform quantization  described in \cite{aysal:SP08}. For alphabet size $L$, the quantization alphabet $\mathcal{Z}$ is defined as the midpoints of $L$ equally-sized quantization bins. The quantizer is defined as
\begin{equation}
	\phi(z) = \min_{q \in \mathcal{Z}} |q - (z+u)|,
\end{equation}
where $u$ is a dither, drawn uniformly and randomly from $[-\Delta/2,\Delta/2)$ each time $\phi$ is called, and $\Delta = 1/L$ is the width of each quantization bin. Statistically, one can write the quantized value as
\begin{equation*}
	\phi(z) = z + v,
\end{equation*}
where $v$ is uniformly distributed across $[-\Delta/2,\Delta/2)$ and independent of $z$.

The alphabet size $L = |\mathcal{Z}|$ depends on the quality of the wireless links. Since we define connectivity at signal-to-noise threshold $\gamma$, we take $L$ to be determined by the Shannon capacity of a wireless link at SNR $\gamma$. Supposing unit bandwidth and slot duration, nodes successfully transmit $K\log_2(1+\gamma)$ bits over the wireless links, which results in an alphabet size of
$	L = \lfloor2^{K\log_2(1+\gamma)}\rfloor = \lfloor (1 + \gamma)^K\rfloor.
$

At time $t$, each node updates its estimate by taking a functional combination of quantized estimates received from neighboring nodes:
\begin{equation}
	z_n(t) = g_n(\phi(\mathbf{z}(t)),t),
\end{equation}
where the function $g_n$ may only depend on estimates $\phi(z_m(t))$ for which $m \in \mathcal{N}_n(t)$. Again the connectivity and  the function $g_n$ may vary with time.

%With quantization it becomes difficult to speak of convergence time. A large class of consensus algorithms does not converge on the true average to within any finite tolerance, precluding our defining $T_\epsilon$ as before. In fact, quantization induces a tradeoff between resource consumption and estimate quality.

For a consensus algorithm with quantization, let $T$ be the number of rounds for which consensus is run. Then let $B$ and $E$ be the time-bandwidth product and total transmit energy, defined as before but with $T$ taking the role of $T_\epsilon$. Finally, define the {\em mean squared error} as
\begin{equation}
	\sigma^2 = \max_{\mathbf{z}(0) \in [0,1)^N} E\left[\frac{1}{N}\sum_{n=1}^N(z_n(T) - z_\mathrm{ave})^2\right],
\end{equation}
where the expectation is taken over any randomness in the quantization operator as well as in the consensus algorithm. There is an inherent tradeoff between the total transmit energy $E$ and the mean-squared error $\sigma^2$. One can always reduce the estimation error by injecting more transmit energy into the network and increasing the rate of the wireless links.

Finally, throughout this paper we rely on the following lemma, which shows that the number of nodes in any region is asymptotically proportional to its area to within an arbitrary tolerance.
\begin{lemma}[Ozgur-Leveque-Tse,\cite{ozgur:IT07}]\label{lem:geo.dense}
	Let $A \subset [0,1] \times [0,1]$ be a region inside the unit square having area $|A|$, and let
	$
		\mathcal{C} = \{n : \mathbf{r}_n \in A\}
	$
	be the nodes lying in $A$. Then, for any $\delta > 0$,
	\begin{equation}
		(1-\delta)|A|N \leq |\mathcal{C}| \leq (1+\delta)|A|N,
	\end{equation}
	with probability greater than $1-1/|A| e^{-\Gamma(\delta)|A|N}$, where $\Gamma(\delta) >0$ is a quantity independent of $N$ and $|A|$.
\end{lemma}

\section{Inner Bounds}\label{sect:inner.bounds}
In this section we derive inner bounds on the resource costs for consensus over the proposed wireless model. We begin with the case of ideal links.
\begin{theorem}\label{thm:lower.bound}
	For any consensus algorithm, and for any $\epsilon = O(1)$, we have, with probability approaching 1 as $N \to \infty$:
	\begin{align}
		T_\epsilon &= B_\epsilon = \Omega(1) \\
		E_\epsilon &= \Omega(N^{1 - \alpha/2}).
	\end{align}
\end{theorem}
\begin{IEEEproof}
	The bounds on $T_\epsilon$ and $B_\epsilon$ are trivial. The bound on $E_\epsilon$ follows from the observation in \cite{benezit:IT10} that consensus to within a constant tolerance requires at least $\Omega(N)$ transmissions. Each transmission from node $n$ consumes at least enough energy to reach its nearest neighbor, which by (\ref{eqn:power.radius}) is equal to $\gamma/G d_\mathrm{min}^\alpha(n)$, where $d_\mathrm{min}(n)$ is the distance between node $n$ and its nearest neighbor. It is well-known (e.g., in \cite{penrose:03}), that $d_{\mathrm{min}}(n) = \Theta(N^{-1/2})$ with high probability, so 
	\begin{equation}
		E_\epsilon \geq \Omega(N) \frac{\gamma}{G} \Theta(N^{-\alpha/2}) \notag
		= \Omega(N^{1-\alpha/2}).
	\end{equation}
\end{IEEEproof}

When links are rate limited, we derive an inner bound on the tradeoff between resources and estimation error.
\begin{theorem}\label{thm:lower.bound.quantized}
	For any consensus algorithm with rate-limited links, any achievable tradeoff in performance metrics satisfies the following with high probability:
	\begin{align}
		T &= B = \Omega(K) \\
		E &= K\sum_{n=1}^N\Omega(N^{u_n-\alpha/2}) \\
		\sigma^2 &= \frac{1}{N} \sum_{n=1}^N \Omega(N^{-2Ku_n}),
	\end{align}
	for $u_n > 0$ and $K > 0$. In particular, choosing each $u_n = u$ yields
	\begin{align}
		E &= \Omega(KN^{1+u-\alpha/2}) \\
		\sigma^2  &= \Omega(N^{-2Ku}).
	\end{align}
\end{theorem}
\begin{IEEEproof}
	As in the ideal-link case, the bounds on $T$ and $B$ are trivial; at least one transmission must be made, which requires $K$ channel uses. To bound the tradeoff between energy and estimation error, momentarily consider a single node $n$. Suppose a genie supplies node $n$ with $z_\mathrm{ave}$, and further suppose that only node $n$'s nearest neighbor, denoted by $m$, needs to compute the average. In this case, the optimal strategy is for $n$ to quantize $z_\mathrm{ave}$ and transmit it directly to node $m$. In principle, other nodes could transmit their measurements to $m$, but since they are no closer order-wise, and since they have only partial knowledge of the average, any energy they expend would be better used by node $n$.
	
	Without loss of generality, let $P_n = N^{u_n-\alpha/2}$ denote the transmit power used by node $n$ to transmit $z_\mathrm{ave}$. Since again the distance between nearest neighbors is $\Theta(N^{-1/2})$, then with high probability the highest possible SNR is $\gamma = \Theta(N^{u_n})$. Therefore, $L = (1+\gamma)^K = \Theta(N^{Ku_n}))$. The square quantization error at node $n$ on $z_\mathrm{ave}$ is $|e_n|^2 = \Theta(L^{-2})=\Theta(N^{-2Ku_n})$. Repeating the argument for each $n$ establishes the result.
\end{IEEEproof}

By contrast with the ideal-link case, with quantization nodes can trade off between resources and quantization error by adjusting both $K$ and the transmit power. In particular, observe that choosing high $K$ reduces the exponent of the estimation error, but only increases the resource consumption by a constant factor. To highlight this tradeoff, we retain $K$ in the scaling laws for quantization, whereas with ideal links we omit $K$.

Also note that the scaling laws are more favorable for higher path-loss exponents, which may give the false impression that increased path-loss {\em decreases} the resources required. So long as $G$ is chosen such that channel gains are less than unity, the absolute resource consumption required for fixed $N$ increases with $\alpha$. The scaling laws merely indicate that increasing the number of nodes---and therefore decreasing the distance between nearest neighbors---is more beneficial for high path-loss. Of course, for sufficiently large $N$ the far-field assumption on which path-loss is based breaks down, at which point the scaling laws no longer hold. For a discussion of this issue in capacity scaling laws, see \cite{franceschetti:IT09,leeSH:IT12}.

\section{Gossip Algorithms}\label{sect:gossip.performance}
In this section we characterize several existing gossip algorithms with respect to the metrics defined in Section \ref{sect:system.model}. We focus on two variants that give a relatively comprehensive look at the state of the art: randomized gossip \cite{boyd:IT06}, which is probably the best-known approach to gossip, and path averaging \cite{benezit:IT10}, which is order optimal in terms of convergence speed\footnote{Due to its similarity with hierarchical averaging, we might suspect that multiscale gossip \cite{tsianos:IT10} has superior performance to the gossip algorithms studied here with respect to our metrics. However, multiscale gossip differs from hierarchical averaging in the crucial sense that it operates using fixed connectivity. As a result, while we do not carry out the analysis here due to space constraints, one can show that the performance of multi-scale gossip is similar to that of path-averaging.}.
% the performance of gossip algorithms with respect to the metrics defined in Section \ref{sect:system.model}. There are, of course, far too many varieties of gossip for us to analyze every scheme, so we have chosen three varieties that we claim give a rather comprehensive look at the gossip landscape. They are {\em randomized gossip} \cite{boyd:IT06}, {\em path-averaging} \cite{benezit:IT10}, and {\em multi-scale gossip} \cite{tsianos:IT10}. Randomized gossip has the benefit of being an early, well-studied contribution from which more recent approaches have been derived, while path-averaging and multi-scale gossip have the benefit of order-optimal performance in terms of number of transmissions.
Our first task is to adapt the graphical nature of gossip to our wireless model. In order to achieve consensus, it is necessary to choose the topology of the network such that the resulting graph is connected, which requires that each node be neighbors with every node within a radius of $\Theta(\sqrt{\log{N}/N})$  \cite{gupta:CDC98}. Therefore the neighborhoods must satisfy, for every node $n$ that transmits during time slot $t$,
\begin{equation*}
	\mathcal{N}_n(t) = \{m : \norm{\mathbf{x}_m - \mathbf{x}_n}_2 < \Theta(\sqrt{\log{N}/N}) \}.
\end{equation*}
By (\ref{eqn:power.radius}), the transmit power is
\begin{equation}
	P_n(t) = \Theta((\log{N}/N)^{\alpha/2}),
\end{equation}
for every node $n$ transmitting during time slot $t$. This holds for both gossip algorithms considered in this section.

\subsection{Randomized Gossip}
We study the {\em synchronized} randomized gossip of \cite{boyd:IT06}. At each time slot $t$, each node is randomly paired up with one of its neighbors. Paired nodes exchange estimates and average the estimates together, which results in the following dynamics:
\begin{equation*}
	\mathbf{z}(t) = \frac{1}{2}(\mathbf{W}(t) + \mathbf{I})\mathbf{z}(t-1),
\end{equation*}
where $\mathbf{W}(t)$ is a randomly-chosen permutation matrix such that $w_{mn} = 1$ only if nodes $m$ and $n$ are neighbors.

In \cite[Theorem 9]{boyd:IT06} the convergence of randomized gossip is characterized. It is shown that the averaging time satisfies
\begin{equation}
	T_\epsilon = \Theta\left(N \frac{\log \epsilon^{-1}}{\log N}\right). \label{eqn:randomized.gossip.convergence.time}
\end{equation}
Using this fact, we derive bounds on the resource consumption of randomized gossip.

\begin{theorem}
	For randomized gossip, the resource consumption scales as follows with high probability:
	\begin{align}
		T_\epsilon &= \Theta\left(N \frac{\log \epsilon^{-1}}{\log N}\right), \\
		B_\epsilon &= \Omega\left(N \log \epsilon^{-1}\right), \\
		E_\epsilon &= \Theta (N^{2-\alpha/2} (\log N)^{\alpha/2-1}  \log \epsilon^{-1}).
	\end{align}
\end{theorem}
\begin{IEEEproof}
	The bound on $T_\epsilon$ follows from (\ref{eqn:randomized.gossip.convergence.time}). Since every node transmits during every time slot $t$, $P_n(t) =\Theta( (\log{N}/N)^{\alpha/2}) \ \forall n,t$. The transmit energy therefore follows
	\begin{align*}
		E_\epsilon &= K\sum_{t=1}^{T_\epsilon} \sum_{n=1}^N \Theta\left( \frac{\log N}{N}\right)^{\alpha/2} \\
		&=  T_\epsilon \Theta(N^{1-\alpha/2} (\log N)^{\alpha/2}) \\
		&=  \Theta(N^{2-\alpha/2} (\log N)^{\alpha/2-1}  \log \epsilon^{-1}).
	\end{align*}
	Next, the required connectivity radius means that each neighborhood is defined by a region of area $\Theta(\log{N}/N)$. By Lemma \ref{lem:geo.dense}, each neighborhood size satisfies
	\begin{align*}
		(1-\delta)\pi \log N \leq |\mathcal{N}_n(t)| &\leq (1+\delta)\pi \log N \\
		 |\mathcal{N}_n(t)| &= \Theta\left( \log N\right)
	\end{align*} 
	with high probability. Plugging this into (\ref{eqn:time.bandwidth}) yields
	\begin{equation*}
		B_\epsilon = KT_\epsilon \Omega\left( \log N \right)
		= \Omega \left( N \log \epsilon^{-1} \right).
	\end{equation*}
\end{IEEEproof}

\subsection{Path Averaging}
Next, we examine {\em path averaging}%, a more sophisticated gossip algorithm proposed in
\cite{benezit:IT10}. Instead of exchanging estimates with a neighbor, in path averaging each node chooses a geographically distant node with which to exchange its estimate; the exchange is facilitated by multi-hop rounding. In addition to facilitating the exchange, the routing nodes add their estimates ``along the way,'' allowing many nodes to average together in a single round. Once the average of all the nodes' estimates is computed at the destination, the result is routed back to the source.

Path averaging is described in an asynchronous framework in which nodes independently ``wake up,'' initiate multi-hop exchanges, and return to idle state sufficiently quickly that no two exchanges overlap in time. Placing path averaging into our synchronous framework, we suppose that at time $t$ a pair of nodes $n,m$ is randomly selected to engage in a multi-hop exchange. Letting $\mathcal{P}(t)$ be the set of nodes routing from $n$ to $m$, we suppose that the $2(|\mathcal{P}(t)|-1)$ transmissions required to route from $n$ to $m$ and back happen sequentially and thus require $2(|\mathcal{P}(t)|-1)$ time slots. At time slot $t+2(|\mathcal{P}(t)|-1)$, a new pair is chosen.
The dynamics for path averaging has the following form:
\begin{equation}
	z_n(t+2(|\mathcal{P}(t)|-1)) =
	\begin{cases}
		\frac{1}{|\mathcal{P}|}\sum_{m \in \mathcal{P}(t)}z_m(t), & n \in \mathcal{P}(t) \\
		z_n(t), & \text{otherwise}
	\end{cases}.
\end{equation}

In \cite[Theorem 2]{benezit:IT10} it is shown that, for a random uniform network,\footnote{Technically, the convergence speed of path averaging is proven over a torus, so the results we prove in the sequel apply to the torus. Later we show numerical results that suggest that the similar results apply to a square network.} the expected path length is $E[|\mathcal{P}(t)|]~=~\Theta(\sqrt{N/\log{N}})$ and the number of exchanges required to achieve $\epsilon$-consensus is $\Theta(\sqrt{N \log{N}}\log{\epsilon^{-1}})$. Combining these facts, the total number of required transmissions is $\Theta(N \log{\epsilon^{-1}})$.

In casting path averaging in our synchronous framework, we have retained the assumption that multi-hop exchanges do not overlap in time. In principle one could construct a synchronous path-averaging gossip in which multiple exchanges occur simultaneously, perhaps reducing the total amount of time required to achieve consensus. In the following theorem, we provide a rather optimistic  bound on the resource consumption of any such synchronous formulation.
\begin{theorem}
	For any synchronous path-averaging gossip, the resource consumption scales as follows with high probability:
	\begin{align}
		T_\epsilon &= B_\epsilon = \Omega\left(\sqrt{\frac{N}{\log N}}\right) \\
		%B_\epsilon = \Omega(\sqrt{\frac{N}{\log N}}) \\
		E_\epsilon &= \Theta(N^{1-\alpha/2}  (\log N)^{\alpha/2}  \log \epsilon^{-1}).
	\end{align}
\end{theorem}
\begin{IEEEproof}
	We prove the bound on $T_\epsilon$ and $B_\epsilon$ by noting that each route has $\Theta(\sqrt{N/\log N})$ hops. Even in the ideal case in which every round of gossip occurs simultaneously, we still require $T_\epsilon =  \Omega(\sqrt{N/\log N})$ sequential transmissions. Supposing that constant bandwidth is sufficient to accommodate the multiple exchanges, the same bound applies to $B_\epsilon$.
	
	To bound $E_\epsilon$ we point out that, as with randomized gossip, $P_n(t) = \Theta((\log{N}/N)^{\alpha/2})$ for every transmission. Since path-averaging requires $\Theta(N \log \epsilon^{-1})$ transmissions, the overall energy consumption scales as
	\begin{equation*}
		E_\epsilon = \Theta(N^{1-\alpha/2} (\log N)^{\alpha/2} \log \epsilon^{-1}).
	\end{equation*}
\end{IEEEproof}

\section{Hierarchical averaging}\label{sect:hierarchical.averaging}
In this section we present {\em hierarchical averaging}. Much like multi-scale gossip \cite{tsianos:IT10} and the hierarchical cooperation of \cite{ozgur:IT07}, in hierarchical averaging we recursively partition the network into geographically defined clusters. Each cluster achieves internal consensus by mutually broadcasting estimates. Nodes within a cluster then cooperatively broadcast their identical estimates to neighboring clusters at the next level. The process continues until the entire network achieves consensus. In the following subsection we describe the recursive partitioning, after which we describe the algorithm in detail and characterize its resource requirements.
 
\subsection{Hierarchical Partitioning}
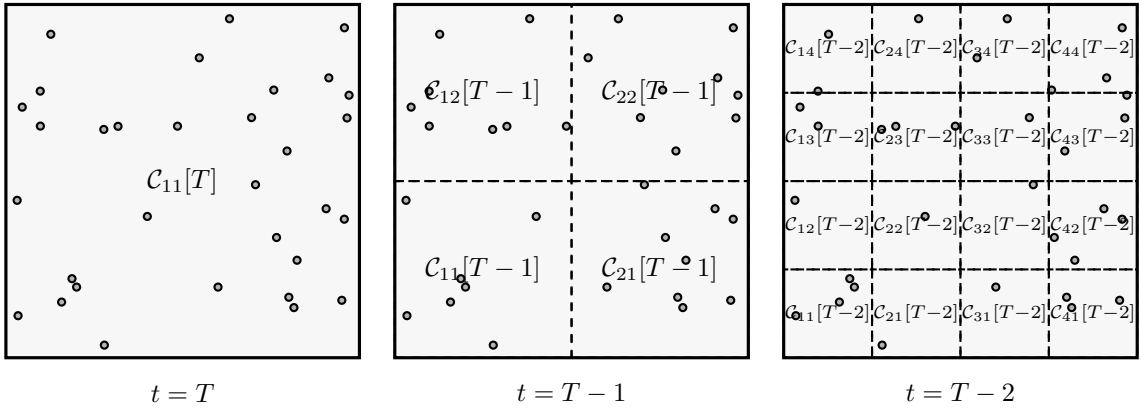
\begin{figure*}[tb]
	\centering
	% !TEX root = ../jsac.tex

\def\radius{0.01cm}
\def\gap{0.01}

\begin{tikzpicture}[scale=4.7,
	terminal/.style={draw=black,fill=white!70!black,thick},
	bottomcluster/.style={very thick,draw=black,fill=white!97!black},
	middlecluster/.style={thick, dashed,draw=black},
	topcluster/.style={thick,dashed,draw=black}
	]
	
	\draw[bottomcluster] (0,0) rectangle (1,1);
	
	\foreach \x/\y in {0.8147/0.1419, 0.9058/0.4218,  0.1270/0.9157, 0.9134/0.7922, 0.6324/0.9595, 0.0975/0.6557, 0.2785/0.0357, 0.5469/0.849, 0.9575/0.9340, 0.9649/0.6787, 0.1576,   0.7577, 0.9706/0.7431, 0.9572/0.3922, 0.4854/0.6555, 0.8003/0.1712, 0.7060/0.4898, 0.0318/0.4456, 0.2769/0.6463, 0.0462/0.7094, 0.0971/0.7547, 0.8235/0.2760, 0.6948/0.6797, 0.3171/0.6551, 0.9502/0.1626, 0.0344/0.1190, 0.7655/0.3404, 0.7952/0.5853, 0.1869/0.2238, 0.2/0.2, 0.4/0.4, 0.6/0.2}
		\draw[terminal] (\x,\y) circle (\radius);
		
		\node at (0.5,0.5) {$\mathcal{C}_{11}[T]$};
		
		\node at (0.5,-0.1) {$t=T$};

	\begin{scope}[xshift = 1.1cm]
		
		\draw[bottomcluster] (0,0) rectangle (1,1);

		\foreach \x/\y in {0.8147/0.1419, 0.9058/0.4218,  0.1270/0.9157, 0.9134/0.7922, 0.6324/0.9595, 0.0975/0.6557, 0.2785/0.0357, 0.5469/0.849, 0.9575/0.9340, 0.9649/0.6787, 0.1576,   0.7577, 0.9706/0.7431, 0.9572/0.3922, 0.4854/0.6555, 0.8003/0.1712, 0.7060/0.4898, 0.0318/0.4456, 0.2769/0.6463, 0.0462/0.7094, 0.0971/0.7547, 0.8235/0.2760, 0.6948/0.6797, 0.3171/0.6551, 0.9502/0.1626, 0.0344/0.1190, 0.3816/0.9597, 0.7655/0.3404, 0.7952/0.5853, 0.1869/0.2238, 0.2/0.2, 0.4/0.4, 0.6/0.2}
		\draw[terminal] (\x,\y) circle (\radius);

		\draw[middlecluster] (0,0) rectangle (0.5,0.5);
		\node at (0.25,0.25) {$\mathcal{C}_{11}[T-1]$};
		\draw[middlecluster] (0,0.5) rectangle (0.5,1);
		\node at (0.25,0.75) {$\mathcal{C}_{12}[T-1]$};
		\draw[middlecluster] (0.5,0) rectangle (1,0.5);
		\node at (0.75,0.25) {$\mathcal{C}_{21}[T-1]$};
		\draw[middlecluster] (0.5,0.5) rectangle (1,1);
		\node at (0.75,0.75) {$\mathcal{C}_{22}[T-1]$};
		
		\node at (0.5,-0.1) {$t=T-1$};

	\end{scope}
	
	\begin{scope}[xshift=2.2cm]
		\draw[bottomcluster] (0,0) rectangle (1,1);
		
		\foreach \x/\y in {0.8147/0.1419, 0.9058/0.4218,  0.1270/0.9157, 0.9134/0.7922, 0.6324/0.9595, 0.0975/0.6557, 0.2785/0.0357, 0.5469/0.849, 0.9575/0.9340, 0.9649/0.6787, 0.1576,   0.7577, 0.9706/0.7431, 0.9572/0.3922, 0.4854/0.6555, 0.8003/0.1712, 0.7060/0.4898, 0.0318/0.4456, 0.2769/0.6463, 0.0462/0.7094, 0.0971/0.7547, 0.8235/0.2760, 0.6948/0.6797, 0.3171/0.6551, 0.9502/0.1626, 0.0344/0.1190, 0.3816/0.9597, 0.7655/0.3404, 0.7952/0.5853, 0.1869/0.2238, 0.2/0.2, 0.4/0.4, 0.6/0.2}
		\draw[terminal] (\x,\y) circle (\radius);
		
		\foreach \i in {1,2,3,4} {
			\foreach \j in {1,2,3,4} {
			\draw[topcluster] (0.25*\i-0.25,0.25*\j-0.25) rectangle (0.25*\i,0.25*\j);
			\node at (0.25*\i - 0.125,0.25*\j - 0.125) {${\scriptstyle \mathcal{C}_{\i\j}[T-2]}$};
		}}
			
		\node at (0.5,-0.1) {$t=T-2$};

	\end{scope}

\end{tikzpicture}
	\caption{Hierarchical partition of the network. Each square cell is divided into four smaller cells, which are each divided into four smaller cells, and so on.}
	\label{fig:recursive.clusters}
\end{figure*}

We partition the network into $T$ sub-network layers, one for each round of consensus, as depicted in Figure \ref{fig:recursive.clusters}. At the top layer, which corresponds to the final round $t=T$ of consensus, there is a single cell. At the next-highest level $t=T-1$, we divide the network into four equal-area square cells. Continuing, we recursively divide each cell into four smaller cells until the lowest layer $t=1$, which corresponds to the first round of consensus. At each level $t$ there are $4^{T-t}$ cells, formally defined as
\begin{multline}
	\mathcal{C}_{jk}(t) = \{ n : \mathbf{r}_n \in [(j-1)2^{t-T},j2^{t-T}) \times \\ [(k-1)2^{t-T},k2^{t-T}) \},
\end{multline}
where $1\leq j,k \leq 2^{T-t}$ index the location of the cell.

Let $\mathcal{C}(n,t)$ denote the unique cluster at layer $t$ containing $n$. Let $M(t) = \sqrt{2}\cdot4^{\frac{t-T}{2}} = \Theta(4^\frac{t-T}{2})$ denote the maximum distance between two nodes in the same cluster at layer $t$.

We choose $T = \lceil \log_4(N^{1-\kappa}) \rceil$, where $0 < \kappa < 1$ is a small constant. In the following lemma we bound the cardinality of each cell.
\begin{lemma}\label{lem:cluster.cardinality}
	For every $1 \leq j,k \leq 2^{T-t}$ and $1 \leq t \leq T$, the cluster cardinalities simultaneously satisfy 
	\begin{equation}
		|\mathcal{C}_{jk}(t)| = \Theta(4^{t}N^{\kappa}),
	\end{equation}
	with probability greater than $1 - N^{2-2\kappa}/16 \cdot e^{-\Gamma(\delta)N^\kappa}$.
\end{lemma}
\begin{IEEEproof}
	By construction, the area of each cell for $t=1$ is
	\begin{align*}
		A &= 4^{1-T} \\
		\implies 4^{1-\log_4(N^{1-\kappa}) - 1} \leq A &\leq 4^{1-\log_4(N^{1-\kappa})} \\
		\frac{N^{\kappa-1}}{16} \leq A &\leq 4N^{\kappa-1}.
	\end{align*}
	By Lemma \ref{lem:geo.dense}, the cardinality of each cell at layer $t=1$ satisfies
	\begin{equation}\label{eqn:cluster.cardinality}
		(1-\delta)\frac{N^{\kappa}}{16} \leq |\mathcal{C}_{jk}(t)| \leq (1+\delta)4 N^{\kappa},
	\end{equation}
	with probability greater than $1 - N^{1-\kappa}/16 \cdot e^{-\Gamma(\delta)N^\kappa}$.
	
	Define $E_{jk}(1)$ as the event in which $|\mathcal{C}_{jk}(1)|$ is outside the bounds specified in (\ref{eqn:cluster.cardinality}). Clearly $\mathrm{Pr}\{E_{jk}(1)\} \leq N^{1-\kappa}/16 \cdot e^{-\Gamma(\delta)N^\kappa}$. By the union bound, the total probability follows
	\begin{align}
		\mathrm{Pr}\left(\bigcup_{1 \leq j,k \leq 2^{T-1}} \!\!\!\!\! E_{j,k}(1)\right) &\leq \sum_{1 \leq j,k \leq 2^{T-1}} N^{1-\kappa}/16 \cdot e^{-\Gamma(\delta)N^\kappa} \notag\\
		&\leq N^{2-2\kappa}/16 \cdot e^{-\Gamma(\delta)N^\kappa} \to 0.
	\end{align}
	Therefore, every cell at $t=1$ simultaneously satisfies $|\mathcal{C}_{jk}(1)| = \Theta(N^{\kappa})$ with the desired probability. Finally, since each cell at layer $t$ is composed of $4^{t-1}$ cells at layer 1, the cardinalities at each layer satisfy the claim with high probability.
\end{IEEEproof}

\subsection{Algorithm Description}\label{sect:cooperative.algorithm}
%Here we lay out the details of hierarchical averaging. We suppose that each node $n$ has the following information about the network: the total number of nodes $N$, its own location $\mathbf{r}_n$, and the number of layers $T$.

First, at time slot $t=1$ each node broadcasts its initial estimate $z_n(0)$ to each member of its cluster $\mathcal{C}(n,t)$. In order to ensure that $n \in \mathcal{N}_m(t)$ for every $m \in \mathcal{C}(n,1)$, each node transmits at power
\begin{align}
	P_n(1) = \frac{\gamma}{G} \max_{m \in \mathcal{C}(n,1)} h_{nm}^\alpha 
	\leq \frac{\gamma}{G} M(1)^\alpha 
	%&\leq \left(2 \cdot 4 N^{\kappa-1}\right)^{\alpha/2} \\
	= O(N^{(\kappa-1)\alpha/2}). \label{eqn:first.round.power}
\end{align}
Each node averages the estimates in its cluster:
\begin{equation}
	z_n(1) = \frac{1}{4^{1-T}N}\sum_{m \in \mathcal{C}(n,1)}z_m(0).
\end{equation}
We use the approximate normalization $1/4^{1-T}N$ instead of the exact factor $1/|\mathcal{C}(n,1)|$ so that nodes at higher levels of the hierarchy need not know the cardinality of the cells. This approximation introduces no error into the final estimate.

After time slot $t=1$, each node in each cluster $\mathcal{C}_{jk}(1)$ has the same estimate, which we denote by $z_{\mathcal{C}_{jk}(1)}(1)$. At each subsequent time slot $2 \leq t \leq T$, each cluster $\mathcal{C}(n,t~-~1)$ cooperatively transmits its estimate to its parent cluster at layer $t$. We take each $P_n(t)$ to be a constant. The transmit power required depends on the phase of the channel gains, as discussed in Section \ref{sect:sub.system.model}. When the phases are fixed and identical, by  (\ref{eqn:coherent.neighborhoods}) the transmit powers must satisfy 
\begin{equation*}
	\left(\sum_{m \in \mathcal{C}(n,t-1)}  h_{mn}P^{1/2}_m(t)\right)^2 = \frac{\gamma}{G},
\end{equation*}
which implies
\begin{align}	
	P_m(t) &= \frac{\gamma}{G\left(\sum_{m \in \mathcal{C}(n,t-1)}h_{mn}\right)^2 } \notag \\
	&\leq \frac{\gamma M(t)^\alpha}{G|\mathcal{C}(n,t-1)|^2} \notag \\
	%&\leq \frac{\gamma (2 \cdot 4^t N^{\kappa-1})^{\alpha/2}}{|\mathcal{C}(n,t-1)|^2} \\
	%&= O\left( \frac{(4^tN^{\kappa-1})^{\alpha/2}}{4^{2t}N^{2\kappa}}\right) \notag \\
	&= O\left(4^{(\alpha/2 - 2)t}N^{-\alpha/2 +\kappa(\alpha/2-2)} \right). \label{eqn:coherent.power}
\end{align}
When the phases are random and uniform, on the other hand, by (\ref{eqn:non.coherent.neighborhoods}) a similar argument shows
\begin{align}
	%\sum_{m \in \mathcal{C}(n,t-1)}  h^2_{mn}P_m(t) &= \frac{\gamma}{G} \notag  \\
	%\implies
	P_m(t) &= \frac{\gamma}{G\sum_{m \in \mathcal{C}(n,t-1)}h^2_{mn} } \notag \\
	&\leq \frac{\gamma M(t)^\alpha}{G|\mathcal{C}(n,t-1)|} \notag \\
	%&\leq \frac{\gamma (2 \cdot 4^t N^{\kappa-1})^{\alpha/2}}{|\mathcal{C}(n,t-1)|} \\
	%&= O\left( \frac{(4^tN^{\kappa-1})^{\alpha/2}}{4^{t-1}}\right) \\
	&= O\left(4^{(\alpha/2 - 1)t}N^{(\kappa-1)\alpha/2} \right). \label{eqn:non.coherent.power}
\end{align} 
After receiving  estimates from the other sub-clusters, each node updates its estimate by taking the sum:
\begin{align}
	z_n(t) &= \frac{1}{4}\sum_{\mathcal{C}(n,t-1) \subset \mathcal{C}(n,t)} z_{\mathcal{C}(n,t-1)} \notag\\
	&= \frac{1}{4^{t-T}N}\sum_{m \in \mathcal{C}(n,t)}z_m(0), \label{eqn:hierarchical.estimate}
\end{align}
where the second equality follows straightforwardly by induction. At time $t$, the identical estimate at each cluster is the average of the measurements from within that cluster.

Consensus is achieved at round $T$, where the four sub-clusters at level $t=T-1$ broadcast their estimates to the entire network. Evaluating (\ref{eqn:hierarchical.estimate}) for $t=T$, we observe that hierarchical averaging achieves perfect consensus; there is no need for a tolerance parameter $\epsilon$. This somewhat surprising result is the consequence of combining the flexibility of wireless, which allows us to adjust the network connectivity at will, with the assumption of ideal links. In the next section we will revisit this assumption.

In the following theorem we derive the resource requirements of hierarchical averaging.
\begin{theorem}\label{thm:phase.coherent}
	With high probability, the resource consumption of hierarchical averaging scales according to
	\begin{align}
		T_\epsilon &= B_\epsilon = O(N^\kappa), \\
		E_\epsilon &=
		\begin{cases}
			O(N^{1-\alpha/2 +\kappa\alpha/2}), & \text{for fixed phase} \\
			O(N^{\kappa\alpha/2}), & \text{for uniform phase}
		\end{cases},
	\end{align}
	for any path-loss exponent $2 \leq \alpha < 4$, for any $\epsilon > 0$ and for any $0 < \kappa \leq 1$.
\end{theorem}
\begin{IEEEproof}
	The bound on $T_\epsilon$ follows by construction; we chose $T = \lceil \log_4 N^{1-\kappa} \rceil = O(N^\kappa)$ layers of hierarchy and constructed the algorithm such that consensus is achieved to within any tolerance $\epsilon>0$,
	
	We derive the bound on $B_\epsilon$ by examining the number of two-hop neighbors for each node. At time slot $t=1$, by (\ref{eqn:first.round.power}) each node transmits at power $P_n(1) = O(N^{(\kappa - 1)\alpha/2})$. The neighbors are nodes within a radius of $O(N^{\kappa-1})$, and the two-hop neighbors are the nodes within twice the radius. Therefore, by Lemma \ref{lem:geo.dense}, the number of two-hop neighbors for each node scales as $O(N^\kappa)$ with probability approaching 1 as $N \to \infty$. By (\ref{eqn:bandwidth.max}), $B(1) = O(N^\kappa)$.
	
	For rounds $2 \leq t \leq T$, we need to bound the number of {\em clusters} within two hops of each node. In (\ref{eqn:coherent.power}) we chose the transmit powers such that the clusters transmit to each node in a circle of area $\pi M^2(t) = O(4^t N^{1-\kappa})$. By construction, each cluster $\mathcal{C}(n,t)$ covers an area of $O(4^tN^{1-\kappa})$. The number of clusters that can fit into the circle---and therefore the number of one-hop neighbors---is a constant. The number of two-hop neighbors, bounded above by the square of the number of one-hop neighbors, also remains constant, so $B(t) = O(1)$. Summing over all rounds, we get
	\begin{equation}
		B_\epsilon = K\sum_{t=1}^T B(t) = O(N^{\kappa}) + \sum_{t=2}^TO(1) = O(N^{\kappa}).
	\end{equation} 	
	
	Finally, we derive the bounds on $E_\epsilon$. For fixed phase, (\ref{eqn:first.round.power}) and (\ref{eqn:coherent.power}) imply that the transmit energy follows
	\small
	\begin{align}
		E_\epsilon %&= K\sum_{t=1}^T \sum_{n=1}^N P_n(t) \notag \\
		&= N\cdot O(N^{(\kappa-1)\alpha/2}) + N\sum_{t=2}^T O\left(4^{(\alpha/2 - 2)t}N^{-\alpha/2 +\kappa(\alpha/2-2)} \right) \notag \\
		%&\leq O(N^{1-\alpha/2 +\kappa\alpha/2}) + O\left(N^{1-\alpha/2+ \kappa(\alpha/2-2)} \sum_{t=0}^{T-1} 4^{(\alpha/2-2)t}\right) \notag\\
		&\stackrel{(a)}{=} O(N^{1-\alpha/2 +\kappa\alpha/2}) + O\left(N^{1-\alpha/2+ \kappa(\alpha/2-2)} \frac{1-4^{(\alpha/2-2)T}}{1-4^{\alpha/2-2}}\right) \notag \\
		&\stackrel{(b)}{=} O(N^{1-\alpha/2 +\kappa\alpha/2}) + O\left(N^{1-\alpha/2+ \kappa(\alpha/2-2)} N^{(\alpha/2-2)(1-\kappa)}\right) \notag\\
		%&= O(N^{1-\alpha/2 +\kappa\alpha/2}) + O\left(N^{-1}\right) \notag \\
		&= O(N^{1-\alpha/2 +\kappa\alpha/2}),
	\end{align}
	\normalsize
	where (a) follows from the finite geometric sum identity, and (b) holds only when $\alpha < 4$. For uniform phase, first note that the condition in (\ref{eqn:non.coherent.power}) is more strict than that of (\ref{eqn:first.round.power}), so the transmit power satisfies
	\begin{equation}\label{eqn:transmit.order.power}
		P_n(t) = O\left(4^{(\alpha/2 - 1)t}N^{(\kappa-1)\alpha/2} \right)
	\end{equation}
	for all $n,t$. Substituting (\ref{eqn:transmit.order.power}) into the definition of $E_\epsilon$ yields
	\begin{align}
		E_\epsilon &= K\sum_{t=1}^T\sum_{n=1}^N P_n(t) \notag\\
		&= O\left(N^{1 + (\kappa-1)\alpha/2}\sum_{t=0}^{T-1} 4^{(\alpha/2-1)t}\right) \notag\\
		%&= O\left(N^{1 + (\kappa-1)\alpha/2}\frac{1-N^{\alpha/2-1}}{1-4^{\alpha/2-1}} \right) \\
		%&= O\left(N^{1 - \alpha/2 + \kappa\alpha/2}(N^{\alpha/2-1} - 1)\right) \\
		&= O(N^{\kappa\alpha/2}),
	\end{align}
	where again we have employed the finite geometric sum identity.
\end{IEEEproof}

Hierarchical averaging achieves resource scaling arbitrarily close to the lower bound of Theorem \ref{thm:lower.bound} when phase is fixed. When phase is uniform, however, the energy consumption is strictly suboptimal for $\alpha>2$. Note that resource scaling does not depend on the channel phases for $\alpha=2$. In this case, hierarchical averaging is order optimal regardless of phase.

\subsection{Numerical Results}
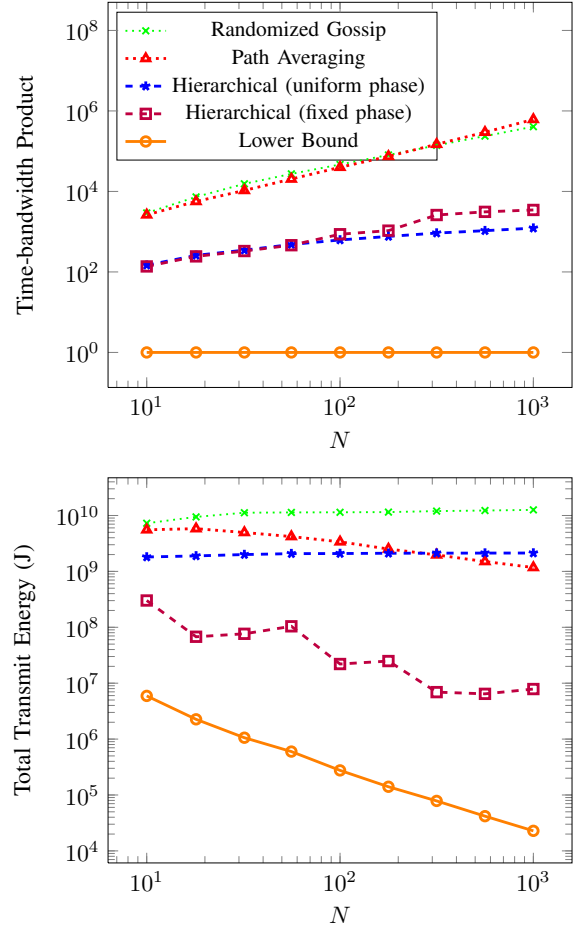
\begin{figure}[htb]
	\centering
	% !TEX root = ../jsac.tex

\begin{tikzpicture}
			
	\begin{scope}[scale=0.9]
	\begin{loglogaxis}[
	xlabel={$N$},
	ylabel={Total Transmit Energy (J)},
	%axis x line=bottom,
	%axis y line=left,
	%height=125pt,
	%width=250pt,
	legend style={at={(0.71,2.21)}}
	]
		
		\addplot[thick,mark=x,mark options=solid,dotted,green] file {jsac/randomE.data};
		\addlegendentry{\small Randomized Gossip};
		
		\addplot[very thick,mark=triangle,mark options=solid,dotted,red] file {jsac/pathAveragingE.data};
		\addlegendentry{\small Path Averaging};
		
		%\addplot[color=green!50!black, thick,dashed] file {jsac/multiscaleE.data};
		%\addlegendentry{\small Multi-scale };
		
		\addplot[very thick,mark=star,mark options=solid,dashed,blue] file {jsac/hierarchicalE.data};
		\addlegendentry{\small Hierarchical (uniform phase)};
		
		\addplot[very thick, mark=square,mark options=solid,dashed,purple] file {jsac/cHierarchicalE.data};
		\addlegendentry{\small Hierarchical (fixed phase)};
		
		\addplot[very thick,mark=o,orange] file {jsac/lowerBoundE.data};
		\addlegendentry{\small Lower Bound };

	\end{loglogaxis}
	\end{scope}
	
	\begin{scope}[scale=0.9,yshift=200]
	\begin{loglogaxis}[
	xlabel={$N$},
	ylabel={Time-bandwidth Product},
	ymax=500000000,
	%axis x line=bottom,
	%axis y line=left,
	%height=125pt,
	%width=250pt,
	%legend style={font=\Large,text=black},
	]

		\addplot[thick,mark=x,mark options=solid,dotted,green]  file {jsac/randomD.data};
		%\addlegendentry{\small Randomized };
		
		\addplot[very thick,mark=triangle,mark options=solid,dotted,red] file {jsac/pathAveragingD.data};
		%\addlegendentry{\small Path Averaging};
		
		%\addplot[color=green!50!black, thick,dashed] file {jsac/multiscaleD.data};
		%\addlegendentry{\small Multi-scale };
		
		\addplot[very thick,mark=star,mark options=solid,dashed,blue] file {jsac/hierarchicalD.data};
		%\addlegendentry{\small Hierarchical  (uniform phase)};
		
		\addplot[very thick, mark=square,mark options=solid,dashed,purple] file {jsac/cHierarchicalD.data};
		%\addlegendentry{\small Hierarchical  (fixed phase)};
		
		\addplot[very thick,mark=o,orange] file {jsac/lowerBoundD.data};

	\end{loglogaxis}
	\end{scope}
\end{tikzpicture}
	\caption{Time-bandwidth product $B_\epsilon$ and transmit energy $E_\epsilon$ as a function of $N$.}
	\label{fig:simulations}
\end{figure}

We examine the empirical performance of the several consensus algorithms presented. We choose $\gamma=10$dB,  $\alpha=4$, $\epsilon=10^{-4}$, $\kappa=10^{-4}$, $K=10$, and $G = 10^{-3\alpha/2}$. We let $N$ run from 10 to 1000, averaging performance over 50 random initializations for each value of $N$. Figure \ref{fig:simulations} shows the average transmit energy $E_\epsilon$ and time-bandwidth product $B_\epsilon$. (Since the data for $T_\epsilon$ are rather similar to that of $B_\epsilon$, we do not plot them.)

%\begin{figure}[htb]
%	\centering
%	\input{jsac/simulations.2.tikz}
%	\caption{Transmit energy $E_\epsilon$ and time-bandwidth product $B_\epsilon$ for a variety of consensus algorithms.}
%	\label{fig:simulations}
%\end{figure}
With respect to time-bandwidth product, hierarchical averaging performs best, the required number of sub-channel uses growing slowly with $N$. The remaining two consensus schemes perform comparably, the required number of sub-channels growing approximately linearly in $N$. Note that, while we bounded the time-bandwidth product of path averaging with a strictly sub-linear term, this bound applied to hypothetical instantiations of the scheme in which multiple transmissions occur simultaneously. Our simulations used the ordinary algorithm, which requires  $\Theta(N)$ sub-channel uses.

With respect to total transmit energy, hierarchical averaging performs best so long as the phases are fixed, in which case the performance is on par order-wise with the lower bound; the oscillations in energy are due to rounding $\log_4(N^{1-\kappa})$ in order to choose $T$. When phases are uniform, however, performance depends on $N$. Even though path averaging has better scaling than hierarchical averaging under uniform phase, for small $N$ hierarchical averaging requires less power. Finally, as expected, randomized gossip requires the most energy in any regime. 

\section{Quantization}\label{sect:quantization}
In this section we examine consensus with quantization. As in the case with ideal links, we first characterize the performance of existing quantized consensus algorithms with respect to the metrics specified in Section \ref{sect:quantization.metrics}. We cannot survey every approach in the literature, so we focus on the {\em quantized consensus} of \cite{kashyap:auto07}, in which consensus is modified to preserve the average of quantized estimates each round. After deriving bounds on its performance, we turn to hierarchical averaging. We show that it achieves the lower bound of Theorem \ref{thm:lower.bound.quantized} when phases are fixed.

\subsection{Quantized Consensus}
In ordinary gossip, the primary difficulty of quantization is that quantizing estimates in general alters the average across the network. Thus, even if consensus is achieved, the dynamics will not in general converge on the true average of the (quantized) measurements. In {\em quantized consensus} \cite{kashyap:auto07}, a family of consensus algorithms is proposed that preserves the average
at each round; it converges to near-consensus around the true average.

Recall from Section \ref{sect:quantization.metrics} that $\mathcal{Z}$ is the set of $L$ points evenly distributed across $[0,1)$, separated by quantization bin width $\Delta = 1/L$. Quantized consensus operates only on quantized values, so first we must quantize the real-valued measurements $z_n(0)$:
\begin{equation}
	q_n(0) = \phi(z_n(0)),
\end{equation}
where $\phi$ is the dithered quantizer described in Section \ref{sect:quantization.metrics}. Let $e_n(0) = \phi(z_n(0)) - z_n(0)$ denote the quantization error.

Much like in randomized gossip, at each round every node randomly selects a neighboring node and mutually averages, with the caveat that one node rounds ``up'' to the nearest member of $\mathcal{Z}$ while the other rounds ``down.'' Letting $i$ and $j$ denote the two nodes in the exchange, we have\footnote{In fact, \cite{kashyap:auto07} proposes a family of algorithms, and the one we use here is only one possibility. The convergence properties we exploit in the following are independent of the specific algorithm chosen.}
\begin{align}
	q_i(t) &= \left\lceil\frac{q_i(t-1) + q_j(t-1)}{2}\right\rceil_{\mathcal{Z}} \\
	q_j(t) &= \left\lfloor\frac{q_i(t-1) + q_j(t-1)}{2}\right\rfloor_{\mathcal{Z}},
\end{align}
where $\lceil \cdot \rceil_{\mathcal{Z}}$ and $\lfloor \cdot \rfloor_\mathcal{Z}$ represent rounding up and down to the nearest element of $\mathcal{Z}$, respectively. In \cite[Theorem 1]{kashyap:auto07} This algorithm is guaranteed to converge on near-consensus: in the limit, each $q_n(t)$ differs by at most a single bin, and the sum of the quantized measurements is preserved. It is difficult to bound the convergence speed of this process in general due to the non-linearity of the updates. However, for the case of a fully-connected graph, in \cite[Lemma 6]{kashyap:auto07} it is shown that quantized consensus requires $\Omega(N^2)$ transmissions over $\Omega(N)$ consensus rounds. Using this fact, we bound the overall performance.
\begin{theorem}\label{thm:quantized.consensus}
	The performance of quantized gossip scales, with high probability, as
	\begin{align}
		T &= \Omega(KN), \\
		B &= \Omega(KN \log N), \\
		E &= \Omega(KN^{2-\alpha/2+u} (\log N)^{\alpha/2}), \text{ and } \\
		\sigma^2 &= \Omega(N^{-2Ku-1}),
	\end{align}
	for any $u\geq 0$.
\end{theorem}
\begin{IEEEproof}
	Choose a target SNR $\gamma$, which may vary with $N$. In order to maintain connectivity, links need to maintain $\gamma$ at radius $\sqrt{\log N/N}$, which requires
	\begin{equation}
		P_n(t) = \Theta\left(\gamma \left(\frac{\log N}{N}\right)^{\alpha/2}\right).
	\end{equation}
	By \cite[Lemma 6]{kashyap:auto07}, consensus requires $\Omega(N)$ rounds for fully-connected graphs, and the performance for random graphs cannot be any better. As in the proof of unquantized randomized gossip, the neighborhood size scales as $\Theta(\log N)$, so the time-bandwidth product scales as
	\begin{equation}
		B = \Omega(K N \log N).
	\end{equation}
	Since consensus requires $\Omega(K N^2)$ total transmissions, the energy scales as
	\begin{equation}
		E = \Omega(K \gamma N^{2-\alpha/2} (\log N)^{\alpha/2}).
	\end{equation}
	
	Finally, we examine the mean-squared error. In the best case, the dynamics converge on true consensus, meaning that $q_n(T)$ is the same for each $n$. In this case the final estimates are merely the average of the quantized measurements $z_n(0)$. Therefore the final estimates are
	\begin{equation*}
		q_n(T) = \frac{1}{N}\sum_{n=1}^N q_n(0)
		%&= \frac{1}{N} \sum_{n=1}^N(z_n(0) - e_n(0)) \\
		= z_\mathrm{ave} - \frac{1}{N}\sum_{n=1}^Ne_n(0),
	\end{equation*}
	where $e_n(0)$ is the quantization error of the initial estimate. In the worst case, each $|e_n(0)| = \Delta/2 = L^{-1}/2$. Since the errors are uncorrelated, the squared error follows
	\begin{align*}
		\sigma^2 &= E\left[ \left(\frac{1}{N}\sum_{n=1}^Ne_n(0)\right)^2\right] \\
		 &\geq \frac{1}{N^2}\sum_{n=1}^NE[|e_n(0)|^2]
		 %&= N^{-1}L^{-2}/2 \\
		= \Omega(N^{-1}L^{-2}).
	\end{align*}
	Recall that $L = \lfloor(1+\gamma)^K\rfloor$. Choosing $\gamma = N^u$ yields the result.
\end{IEEEproof}

We hasten to point out that the bounds here are rather generous, since we supposed that convergence on a random graph is as fast as on a fully-connected graph. In practice, as we shall see in the numerical results presented later, the performance is somewhat worse.

\subsection{Hierarchical Averaging}
We characterize the performance of hierarchical averaging with quantization. As before, cells of nodes at lower layers achieve local consensus, after which they broadcast their estimates to nearby clusters, continuing the process until global consensus is achieved. Here, however, each estimate is quantized prior to transmission, which introduces error that accumulates during consensus.

%As in the previous subsection, we employ the dithered quantizer. For the uniform quantization alphabet with cardinality $L$, let the quantized version of the estimate $z_n(t)$ be denoted
%\begin{equation}
%	q_n(t) = \phi(z_n(t)).
%\end{equation}
%We can write each quantized value as
%\begin{equation}
%	q_n(t) = z_n(t) + v_n(t),
%\end{equation}
%where each $v_n(t)$ is uniform over $[-\Delta/2,\Delta,2)$ and independent for every $n,t$.

We choose $T=\lceil \log_4 N^{1-\kappa} \rceil$ and define the cells $\mathcal{C}_{jk}(t)$ as before. At time slot $t=1$, each node $n$ quantizes its initial measurement $z_n(0)$ and broadcasts the quantized value to the nodes in $\mathcal{C}(n,1)$. Following (\ref{eqn:first.round.power}), this requires
\begin{equation}
	P_n(1) = O(\gamma N^{(\kappa-1)\alpha/2}),
\end{equation}
Each node $n$ updates its estimate by averaging the quantized estimates in its cluster:
\begin{align}
	z_n(1) &= %\frac{1}{4^{1-T}N}\sum_{m \in \mathcal{C}(n,1)}q_m(0) \\
	\frac{1}{4^{1-T}N}\sum_{m \in \mathcal{C}(n,1)}z_m(0) + v_m(0).
\end{align}
As before we use the normalization factor $1/4^{1-T}N$ in order to avoid nodes' needing to know the cluster cardinalities. Next, at time slot $2 \leq t \leq T$, each cluster at layer $t-1$ quantizes its estimate and cooperatively broadcasts to the members of its parent cluster at layer $t$. Following (\ref{eqn:coherent.power}) and (\ref{eqn:non.coherent.power}), this requires
\begin{equation}
	P_n(t) = 
	\begin{cases}
		O(\gamma4^{(\alpha/2-2)t}N^{-\alpha/2+\kappa(\alpha/2-2)}), & \text{fixed phases} \\
		O(\gamma4^{(\alpha/2-1)t}N^{(\kappa-1)\alpha/2}), &\text{uniform phases}
	\end{cases}.
\end{equation}

	At time step $t>1$, each node $n$ averages together the estimates from each of the subclusters $\mathcal{C}(m,t-1) \subset \mathcal{C}(n,t)$. By induction, this estimate is
%	yielding
%	\begin{align*}
%		z_n(2) &= \frac{1}{4}\sum_{\mathcal{C}(m,1) \subset \mathcal{C}(n,2)} q_{\mathcal{C}(m,1)}(1) \\ %\displaybreak
%		&= \frac{1}{4}\sum_{\mathcal{C}(m,1) \subset \mathcal{C}(n,2)}z_{\mathcal{C}(m,1)}(1) + v_{\mathcal{C}(m,1)}(1) \\ 
%		&= \frac{1}{4}\sum_{\mathcal{C}(m,1) \subset \mathcal{C}(n,2)} \left(\frac{1}{4^{1-T}N}\sum_{k \in \mathcal{C}(m,1)}z_k(0) + v_k(0)\right) + v_m(1) \\
%		&= \frac{1}{4^{2-T}N}\sum_{k \in \mathcal{C}(n,2)} (z_k(0) + v_k(0)) + \frac{1}{4}\sum_{\mathcal{C}(m,1) \subset \mathcal{C}(n,2)} v_{\mathcal{C}(m,1)}(1).
%	\end{align*}
%	Continuing by induction, at arbitrary round $t$ the estimate is
	\begin{multline}\label{eqn:quantized.estimates}
		z_n(t) = \frac{1}{4^{t-T}N}\sum_{k \in \mathcal{C}_{(n,t)}}(z_k(0) + v_k(0)) \\ + \sum_{s=1}^{t-1} \sum_{M \in R_n(t,s)} 4^{s-t} v_M(s),
	\end{multline}
	where $R_n(t,s)$ is the set of all clusters $\mathcal{C}(m,s)$ that are subsets of $\mathcal{C}(n,t)$.  In other words, at round $t$ we have the total average so far, corrupted by quantization noise from each of the rounds $s < t$.
In the following theorem, we detail the resource-estimate tradeoff achieved by this scheme.
\begin{theorem}
	Using dithered quantization, hierarchical averaging achieves the following tradeoff between resource consumption and estimation error with high probability:
	\begin{align}
		T &= B = O(KN^\kappa) \\
		E &=
		\begin{cases}
			O(KN^{1-\alpha/2+\kappa\alpha/2 + u}), & \text{for fixed phases} \\
			O(KN^{\kappa\alpha/2 + u}), &\text{for uniform phases}
		\end{cases} \\
		\sigma^2 &= \Theta(N^{-2Ku}),
	\end{align}
	for any $u \geq 0$, $\kappa > 0$, and $2 \leq \alpha < 4$. In particular, for $u=0$ the estimation error is constant in the network size using the same amount of energy as in the non-quantized case.
\end{theorem}
\begin{IEEEproof}
	Choose an SNR $\gamma$, which may vary with $N$. Since the number of rounds and the cluster geometry is unchanged from the non-quantized case, we can repeat the argument from Theorem \ref{thm:phase.coherent}, yielding $T = B = O(KN^{\kappa})$. Repeating the arguments from Theorem \ref{thm:phase.coherent}, which yields
	\begin{equation*}
		E =
		\begin{cases}
			O(\gamma N^{1-\alpha/2+\kappa\alpha/2}), & \text{for fixed phases} \\
			O(\gamma N^{\kappa\alpha/2}), &\text{for uniform phases}
		\end{cases}.
	\end{equation*}
	
	All that remains is to bound the estimation error. Evaluating (\ref{eqn:quantized.estimates}) for $t=T$, we get, for every $n$
	\begin{align*}
		z_n(T) &= \frac{1}{N}\sum_{k =1}^N(z_k(0) + v_k(0)) + \sum_{s=1}^{T-1} \sum_{M \in R_n(t,s)} 4^{s-T} v_M(s) \\
		&= z_{\mathrm{ave}} +  \sum_{s=0}^{T-1} \sum_{M \in R_n(t,s)} 4^{s-T} v_M(s).
	\end{align*}
	The mean squared estimation error is therefore
	\begin{align*}
		\sigma^2 &= E\left[\left|\sum_{s=0}^{T-1} \sum_{M \in R_n(t,s)} 4^{s-T} v_M(s)\right|^2\right] \\
		&=  \sum_{s=0}^{T-1} \sum_{M \in R_n(t,s)} 4^{2(s-T)} E[|v_M(s)|^2],
	\end{align*}
	where the equality is due to the independence of the quantization error terms. Since each $v_M(s)$ is uniformly distributed across $[-\Delta,\Delta)$, we have $E[|v_m(2)|^2] = \Theta(\Delta^2) = \Theta(L^{-2})$. Therefore, we have
	\begin{align*}
		\sigma^2 &= \Theta(L^{-2})\sum_{s=0}^{T-1} \sum_{M \in R_n(t,s)} 4^{2(s-T)} \\
		&= \Theta(L^{-2})\sum_{s=0}^{T-1} 4^{T-s} 4^{2(s-T)} \\
		%&= \Theta(L^{-2})\sum_{s=0}^{T-1} 4^{s-T} \\
		%&= \Theta(L^{-2}N^{-1}) \sum_{s=0}^{T-1}4^s \\
		&= \Theta(L^{-2}N^{-1}) \frac{1-4^T}{1-4}
		= \Theta(L^{-2}),
	\end{align*}
	since $4^T = \Theta(N)$. Recalling that $L = \Theta(\gamma^K)$, and choosing $\gamma = N^u$, yields the result.
\end{IEEEproof}

\subsection{Numerical Results}
\begin{figure}[htb]
	\centering
	% !TEX root = ../jsac.tex

\begin{tikzpicture}
			
	\begin{scope}[scale=0.9]
	\begin{loglogaxis}[
	xlabel={$N$},
	ylabel={Total Transmit Energy (J)},
	%axis x line=bottom,
	%axis y line=left,
	%height=125pt,
	%width=250pt,
	%legend style={at={(1.05,1.05)}}
	]

		\addplot[very thick,mark=x,mark options=solid,dotted,color=green] file{jsac/randomE.q.data};
		
		\addplot[very thick,mark=o,mark options=solid,dotted,color=red] file {jsac/quantizedE.q.data};	
		
		\addplot[very thick,mark=star,mark options=solid,dashed,color=blue] file {jsac/hierarchicalE.q.data};

		\addplot[very thick, mark=square,mark options=solid,dashed,color=purple] file {jsac/cHierarchicalE.q.data};

	\end{loglogaxis}
	\end{scope}
	
	\begin{scope}[scale=0.9,yshift=200]
	\begin{loglogaxis}[
	xlabel={$N$},
	ylabel={Mean-square Error},
	ymax=0.003,
	%axis x line=bottom,
	%axis y line=left,
	%height=125pt,
	%width=250pt,
	%legend style={at={(.45,1.25)}}
	]

		\addplot[very thick,mark=x,mark options=solid,dotted,color=green] file {jsac/randomMSE.q.data};
		\addlegendentry{\small Randomized};

		\addplot[very thick,mark=o,mark options=solid,dotted,color=red] file {jsac/quantizedMSE.q.data};
		\addlegendentry{\small Quantized };
		
		%\addplot[color=red, thick,dotted] file {jsac/pathAveragingMSE.q.data};
		%\addlegendentry{\small Path Averaging};
		
		%\addplot[color=green!50!black, thick,dashed] file {jsac/multiscaleD.data};
		%\addlegendentry{\small Multi-scale };
		
		\addplot[very thick,mark=star,mark options=solid,dashed,color=blue] file {jsac/hierarchicalMSE.q.data};
		\addlegendentry{\small Hierarchical  (uniform phase)};
		
		\addplot[very thick, mark=square,mark options=solid,dashed,color=purple] file {jsac/cHierarchicalMSE.q.data};
		\addlegendentry{\small Hierarchical  (fixed phase)};

	\end{loglogaxis}
	\end{scope}
\end{tikzpicture}
	\caption{Mean-square error $\sigma^2$ and total energy $E$ as a function of $N$.}
	\label{fig:q.simulations}
\end{figure}
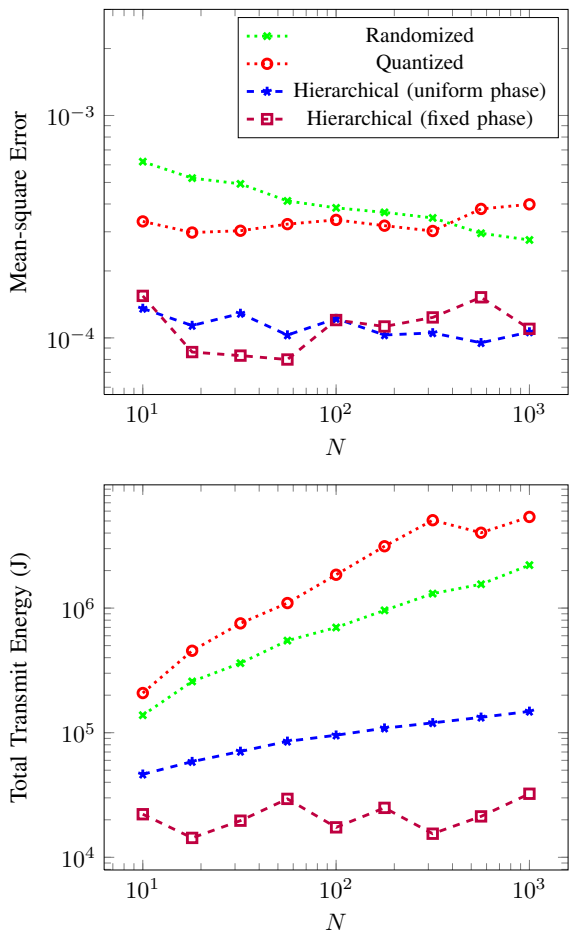

We examine the empirical performance of the quantized consensus discussed. We also run simulations for randomized gossip, employing dithered quantization to accommodate the finite-rate links. We again choose $\gamma=10$dB, $\kappa=10^{-4}$, $K=10$, and $G = 10^{-3\alpha/2}$, and we again let $N$ run from 10 to 1000 and average performance over 50 initializations, but here we choose $\alpha=2$. Choosing $\gamma$ constant means that the quantization error $\Delta$ is constant in $N$, and the minimum quantization error is itself constant. In Figure \ref{fig:q.simulations} we plot the energy $E$ against the mean-square error $\sigma^2$.

The energy expenditure for hierarchical averaging is consistent with theory, although we note that uniform phase results in higher expenditure than fixed phase, even though the scaling laws are the same. This is due to the coherence gain obtained when phases are fixed. The energy expenditure for randomized gossip increases roughly linearly in $N$, suggesting that the energy burden with fixed $\gamma$ is similar to the non-quantized case. As expected, quantized consensus performs worse than predicted by Theorem \ref{thm:quantized.consensus} in terms of energy consumption. Since $\gamma$ is chosen as a constant, the mean-squared error remains approximately constant for each scheme.

\section{Conclusion}\label{sect:conclusion}
We have studied consensus from an explicitly wireless perspective, defining a realistic propagation environment, confronting interference, and characterizing the required resource consumption in terms of energy, time and bandwidth. For dense networks and under path-loss propagation, we have shown that consensus is {\em cheap}. In the worst case, the resource costs for consensus are nearly constant in the size of the network. Furthermore, we have shown that quantization is only a minor issue in consensus. By increasing the block length of transmissions, one can force the quantization error to decay as quickly as desired while paying only a constant penalty in energy, time, and bandwidth. We have shown that these gains can be realized with practical algorithms, such as hierarchical averaging, which is nearly order-optimal with respect to resource consumption for many environments.

%In this paper we have studied consensus from an explicitly wireless perspective, defining a path-loss model, confronting interference, and defining resource consumption in terms of energy, time, and bandwidth. We have shown that, while existing consensus algorithms such as gossip may be order-optimal with respect to the amount of energy required to achieve consensus, they are strictly suboptimal with respect to the time and bandwidth required. Additionally, we have proposed {\em hierarchical averaging}, an approach to consensus derived explicitly for wireless. For free-space propagation, or for any $2 \leq \alpha < 4$ and with fixed channel phase, hierarchical averaging is nearly order optimal with respect to all three metrics simultaneously. We also examined the effects of quantization. Using dithered quantization, we showed that, without expending any additional energy over the non-quantized case, hierarchical averaging suffers only from bounded estimation error in the size of the network. Therefore, hierarchical averaging appears to be an efficient, robust approach to consensus over wireless networks.

%Nevertheless, our results are rather preliminary.
Nevertheless, our wireless model involves several simplifying assumptions. We supposed synchronous transmission, that out-of-range nodes do not contribute interference, and that channel gains are determined entirely by path-loss. Future work includes exploring the effects of relaxing these assumptions, incorporating medium-access techniques such as CSMA, or characterizing performance in the presence of channel fades. We expect that, since it is cooperative in nature, hierarchical averaging is particularly robust to fading.

%However, in constructing a wireless model in which analysis is tractable, we made several simplifying assumptions. In particular, we supposed that nodes are synchronized, that out-of-range nodes do not interfere with other nodes, and that channel gains are completely determined by the path-loss model. Future work involves exploring the effect of relaxing these assumptions. In the case of synchronization, one could incorporate the costs of medium-access techniques such as CSMA. In the case of interference, one could model neighborhoods according to a signal-to-{\em interference}-plus-noise ratio and derive optimal scaling laws. In the case of channel gains, the effects of fading and outage merit study. We expect that, since it is cooperative in nature, hierarchical averaging is particularly robust to outage.

Finally, improvements to hierarchical averaging are possible. In the case of uniform phase, hierarchical averaging was suboptimal with respect to transmit energy for $\alpha > 2$, since in this case transmissions to not coherently combine at receivers. In \cite{ozgur:IT07}, distributed {\em receiver} cooperation is used to overcome this difficulty and secure order-optimal performance. We expect that a similar approach will suffice for consensus, regardless of channel phases or path-loss exponents. 

\bibliography{/Users/nokleby/documents/LaTeX/bibliography}

\begin{thebibliography}{10}
\providecommand{\url}[1]{#1}
\csname url@samestyle\endcsname
\providecommand{\newblock}{\relax}
\providecommand{\bibinfo}[2]{#2}
\providecommand{\BIBentrySTDinterwordspacing}{\spaceskip=0pt\relax}
\providecommand{\BIBentryALTinterwordstretchfactor}{4}
\providecommand{\BIBentryALTinterwordspacing}{\spaceskip=\fontdimen2\font plus
\BIBentryALTinterwordstretchfactor\fontdimen3\font minus
  \fontdimen4\font\relax}
\providecommand{\BIBforeignlanguage}[2]{{%
\expandafter\ifx\csname l@#1\endcsname\relax
\typeout{** WARNING: IEEEtran.bst: No hyphenation pattern has been}%
\typeout{** loaded for the language `#1'. Using the pattern for}%
\typeout{** the default language instead.}%
\else
\language=\csname l@#1\endcsname
\fi
#2}}
\providecommand{\BIBdecl}{\relax}
\BIBdecl

\bibitem{cybenko:JPDC89}
\BIBentryALTinterwordspacing
G.~Cybenko, ``Dynamic load balancing for distributed memory multiprocessors,''
  \emph{Journal of Parallel and Distributed Computing}, vol.~7, no.~2, pp. 279
  -- 301, 1989. [Online]. Available:
  \url{http://www.sciencedirect.com/science/article/pii/074373158990021X}
\BIBentrySTDinterwordspacing

\bibitem{tsitsiklis:PHD84}
J.~N. Tsitsiklis, ``Problems in decentralized decision making and
  computation,'' Ph.D. dissertation, Massachusets Institute of Technology,
  Cambridge, MA, Nov. 1984.

\bibitem{tsitsiklis:TAC86}
J.~Tsitsiklis, D.~Bertsekas, and M.~Athans, ``Distributed asynchronous
  deterministic and stochastic gradient optimization algorithms,'' \emph{{IEEE}
  Trans. Automatic Control}, vol.~31, no.~9, pp. 803 -- 812, Sept. 1986.

\bibitem{duchi:TAC12}
J.~Duchi, A.~Agarwal, and M.~Wainwright, ``Dual averaging for distributed
  optimization: Convergence analysis and network scaling,'' \emph{{IEEE} Trans.
  Automatic Control}, vol.~57, no.~3, pp. 592 --606, Mar. 2012.

\bibitem{saligrama:TSP06}
V.~Saligrama, M.~Alanyali, and O.~Savas, ``Distributed detection in sensor
  networks with packet loss and finite capacity links,'' \emph{{IEEE} Trans.
  Signal Processing}, vol.~54, no.~11, pp. 4118--4132, Nov. 2006.

\bibitem{spanos:ISIPSN}
\BIBentryALTinterwordspacing
D.~P. Spanos, R.~Olfati-Saber, and R.~M. Murray, ``Approximate distributed
  {K}alman filtering in sensor networks with quantifiable performance,'' in
  \emph{Proceedings of the 4th International Symposium on Information
  Processing in Sensor Networks}, 2005. [Online]. Available:
  \url{http://dl.acm.org/citation.cfm?id=1147685.1147711}
\BIBentrySTDinterwordspacing

\bibitem{boyd:IT06}
S.~Boyd, A.~Ghosh, B.~Prabhakar, and D.~Shah, ``Randomized gossip algorithms,''
  \emph{{IEEE} Trans. Info. Theory}, vol.~52, no.~6, pp. 2508 -- 2503, June
  2006.

\bibitem{benezit:IT10}
F.~Benezit, A.~Dimakis, P.~Thiran, and M.~Vetterli, ``Order-optimal consensus
  through randomized path averaging,'' \emph{Information Theory, IEEE
  Transactions on}, vol.~56, no.~10, pp. 5150 --5167, oct. 2010.

\bibitem{tsianos:IT10}
K.~I. Tsianos and M.~G. Rabbat, ``Multiscale gossip for efficient decentralized
  averaging in wireless packet networks,'' \emph{submitted to {IEEE} Trans.
  Info. Theory}, 2010.

\bibitem{gupta:CDC98}
P.~Gupta and P.~Kumar, ``Critical power for asymptotic connectivity,'' in
  \emph{Proc. Conference on Decision and Control}, vol.~1, 1998, pp. 1106
  --1110.

\bibitem{ozgur:IT07}
A.~Ozgur, O.~Leveque, and D.~Tse, ``Hierarchical cooperation achieves optimal
  capacity scaling in ad hoc networks,'' \emph{{IEEE} Trans. Info. Theory},
  vol.~53, no.~10, pp. 3549 --3572, Oct. 2007.

\bibitem{dimakis:TSP08}
A.~D.~G. Dimakis, A.~D. Sarwate, and M.~J. Wainwright, ``Geographic gossip:
  Efficient averaging for sensor networks,'' \emph{IEEE Trans. Signal
  Processing}, vol.~56, no.~3, pp. 1205--1216, March 2008.

\bibitem{ustebay:Allerton08}
D.~Ustebay, B.~Oreshkin, M.~Coates, and M.~Rabbat, ``Rates of convergence for
  greedy gossip with eavesdropping,'' in \emph{Proc. Allerton}, Monticello, IL,
  2008.

\bibitem{aysal:TSP09}
T.~C. Aysal, M.~E. Yildiz, A.~D. Sarwate, and A.~Scaglione, ``Broadcast gossip
  algorithms for consensus,'' \emph{{IEEE} Trans. Signal Processing}, vol.~57,
  no.~7, pp. 2748--2761, July 2009.

\bibitem{nazer:STSP11}
B.~Nazer, A.~G. Dimakis, and M.~Gastpar, ``Local interference can accelereate
  gossip algorithms,'' \emph{IEEE J. Selected Topics Signal Proc.}, vol.~5,
  no.~4, pp. 876--887, Aug. 2011.

\bibitem{sardellitti:SP12}
S.~Sardellitti, S.~Barbarossa, and A.~Swami, ``Optimal topology control and
  power allocation for minimum energy consumption in consensus networks,''
  \emph{{IEEE} Trans. Signal Processing}, vol.~60, no.~1, pp. 383--399, 2012.

\bibitem{vanka:JSCC10}
S.~Vanka, V.~Gupta, and M.~Haenggi, ``Power-delay analysis of consensus
  algorithms on wireless networks with interference,'' \emph{Int. J. Systems,
  Control, and Communications}, vol.~2, no.~3, pp. 256--274, 2010.

\bibitem{vanka:JSTSP11}
S.~Vanka, M.~Haenggi, and V.~Gupta, ``Convergence speed of the consensus
  algorithm with interference and sparse long-range connectivity,'' \emph{IEEE
  J. Selected Topics Signal Proc.}, vol.~5, no.~4, pp. 855--865, Aug. 2011.

\bibitem{xiao:JPDC07}
L.~Xiao, S.~Boyd, and S.-J. Kim, ``Distributed average consensus with
  least-mean-square deviation,'' \emph{Journal of Parallel and Distributed
  Computing}, vol.~67, no.~1, pp. 33 -- 46, 2007.

\bibitem{kashyap:auto07}
A.~Kashyap, T.~Basar, and R.~Srikant, ``Quantized consensus,''
  \emph{Automatica}, vol.~43, no.~7, pp. 1192 -- 1203, 2007.

\bibitem{benezit:JSAC11}
F.~Benezit, P.~Thiran, and M.~Vetterli, ``The distributed multiple voting
  problem,'' \emph{IEEE J. Selected Topics Signal Proc.}, vol.~5, no.~4, pp.
  791 --804, Aug. 2011.

\bibitem{aysal:SP08}
T.~Aysal, M.~Coates, and M.~Rabbat, ``Distributed average consensus with
  dithered quantization,'' \emph{Signal Processing, IEEE Transactions on},
  vol.~56, no.~10, pp. 4905 --4918, oct. 2008.

\bibitem{yildiz:SP08}
M.~Yildiz and A.~Scaglione, ``Coding with side information for rate-constrained
  consensus,'' \emph{{IEEE} Trans. Signal Processing}, vol.~56, no.~8, pp. 3753
  --3764, Aug. 2008.

\bibitem{brooks:CPS41}
R.~L. Brooks, ``On colouring the nodes of a network,'' \emph{Mathematical
  Proceedings of the Cambridge Philosophical Society}, vol.~37, pp. 194--197,
  1941.

\bibitem{xiao:SCL04}
L.~Xiao and S.~Boyd, ``Fast linear iterations for distributed averaging,''
  \emph{Systems and Control Letters}, vol.~53, no.~1, pp. 65--78, Sept. 2004.

\bibitem{penrose:03}
M.~Penrose, \emph{Random geometric graphs}.\hskip 1em plus 0.5em minus
  0.4em\relax Oxford, UK: Oxford University Press, 2003, vol.~5.

\bibitem{franceschetti:IT09}
M.~Franceschetti, M.~Migliore, and P.~Minero, ``The capacity of wireless
  networks: Information-theoretic and physical limits,'' \emph{{IEEE} Trans.
  Info. Theory}, vol.~55, no.~8, pp. 3413 --3424, Aug. 2009.

\bibitem{leeSH:IT12}
S.~Lee and S.~Chung, ``Capacity scaling of wireless ad hoc networks: {S}hannon
  meets {M}axwell,'' \emph{{IEEE} Trans. Info. Theory}, vol.~58, no.~3, pp.
  1702--1715, 2012.

\end{thebibliography}
\vspace{-10 mm}
\begin{IEEEbiography}[{\includegraphics[width=1in,height=1.25in,clip,keepaspectratio]{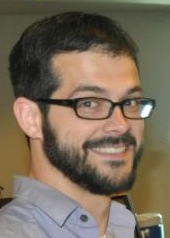}}]{Matthew Nokleby}
(S'04--M'13) received the B.S. {\em (cum laude)} and M.S. degrees from Brigham Young University, Provo, UT, in 2006 and 2008, and the Ph.D. degree from Rice University in 2012. He is currently a Research Scientist at Duke University, Durham, NC. He received a Texas Instruments Distinguished Fellowship from 2008--2012. His research interests include cooperative communications, network information theory, inference over wireless sensor networks, and information theory for high-dimensional signal processing.
\end{IEEEbiography}
\vspace{-7 mm}
\begin{IEEEbiography}[{\includegraphics[width=1in,height=1.25in,clip,keepaspectratio]{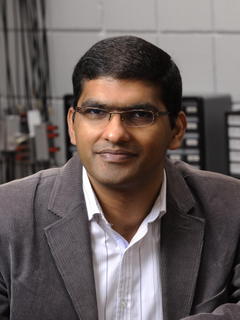}}]{Waheed U. Bajwa}
(S'98--M'09)
%is currently an Assistant Professor in the Department of Electrical and Computer Engineering, Rutgers University, New Brunswick, NJ. His research interests include high-dimensional inference, mathematical signal processing, and complex networks. His detailed biography is available in Feb. 1, 2013 issue of IEEE Trans. Signal Processing.
received the B.E. degree (with Honors) in electrical engineering from the National University of Sciences and Technology,Pakistan, in 2001, and the M.S. and Ph.D. degrees in electrical engineering from the University of Wisconsin-Madison in 2005 and 2009, respectively. He was a Postdoctoral Research Associate in the Program in Applied and Computational Mathematics, Princeton University, Princeton, NJ, from 2009 to 2010, and a Research Scientist in the Department of Electrical and Computer Engineering, Duke University, Durham, NC, from 2010 to 2011. He is currently an Assistant Professor in the Department of Electrical and Computer Engineering, Rutgers University, New Brunswick, NJ. His research interests include high-dimensional inference and inverse problems, compressed sensing, statistical signal processing, wireless communications, and applications in biological sciences, complex networked systems, and radar and image processing. He has more than three years of industry experience, including a summer position with GE Global Research in 2006.

Dr. Bajwa received the Best in Academics Gold Medal and PresidentÕs Gold Medal in Electrical Engineering from the National University of Sciences and Technology (NUST) in 2001, and the Morgridge Distinguished Graduate Fellowship from the University of Wisconsin-Madison in 2003. He was Junior NUST Student of the Year (2000), Wisconsin Union Poker Series Champion (Spring 2008), President of the University of Wisconsin-Madison chapter of Golden Key International Honor Society (2009), and Rutgers UniversityÕs nominee for the Packard Fellowship for Science and Engineering Award (2012). He served as a Guest Associate Editor for a special issue on ``Compressive Sensing in Communications'' of ElsevierÕs Physical Communication Journal (2010Ð2011). He is currently a member of Golden Key International Honor Society and a co-organizer of the CPSWeek 2013/IPSN Workshop on ``Signal Processing Advances in Sensor Networks.''
\end{IEEEbiography}
%\vspace{-30 mm}
\begin{IEEEbiography}[{\includegraphics[width=1in,height=1.25in,clip,keepaspectratio]{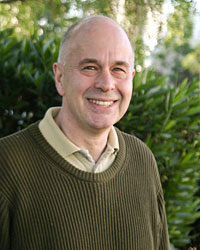}}]{Robert Calderbank}
Robert Calderbank (M'89--SM'97--F'98)
%received the BSc degree in 1975 from Warwick University, England, the MSc degree in 1976 from Oxford University, England, and the PhD degree in 1980 from the California Institute of Technology, all in mathematics. He is Dean of Natural Sciences at Duke University, and was previously Professor of Electrical Engineering and Mathematics at Princeton University where he directed the Program in Applied and Computational Mathematics. Prior to joining Princeton in 2004, he was Vice President for Research at AT\&T. Dr. Calderbank was elected to the US National Academy of Engineering in 2005.
received the B.Sc. degree in 1975 from Warwick University, England, the M.Sc. degree in 1976 from Oxford University, England, and the Ph.D. degree in 1980 from the California Institute of Technology, Pasadena, all in mathematics. He is Dean of Natural Sciences at Duke University, Raleigh, NC. He was previously a Professor of Electrical Engineering and Mathematics, Princeton University, Princeton, NJ, where he directed the Program in Applied and Computational Mathematics. Prior to joining Princeton University in 2004, he was Vice President for Research at AT\&T, responsible for directing the first industrial research lab in the world where the primary focus is data at scale. At the start of his career at Bell Labs, his innovations were incorporated in a progression of voiceband modem standards that moved communications practice close to the Shannon limit. Together with Peter Shor and colleagues at AT\&T Labs, he showed that good quantum error correcting codes exist and developed the group theoretic framework for quantum error correction. He is a co-inventor of space-time codes for wireless communication, where correlation of signals across different transmit antennas is the key to reliable transmission.

Dr. Calderbank served as Editor-in-Chief of the IEEE TRANSACTIONS ON INFORMATION THEORY from 1995 to 1998, and as Associate Editor for Coding Techniques from 1986 to 1989. He was a member of the Board of Governors of the IEEE Information Theory Society from 1991 to 1996 and from 2006 to 2008. He was honored by the IEEE Information Theory Prize Paper Award in 1995 for his work on the Z4 linearity of Kerdock and Preparata Codes (joint with A. R. Hammons Jr., P. V. Kumar, N. J. A. Sloane, and P. Sole), and again in 1999 for the invention of space-time codes (with V. Tarokh and N. Seshadri). He received the 2006 IEEE Donald G. Fink Prize Paper Award and the IEEE Millennium Medal, and was elected to the U.S. National Academy of Engineering in 2005.
\end{IEEEbiography}
%\vspace{-30 mm}
\begin{IEEEbiography}[{\includegraphics[width=1in,height=1.25in,clip,keepaspectratio]{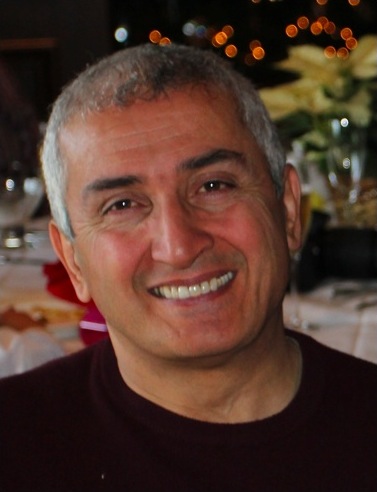}}]{Behnaam Aazhang}
(S'82--M'82--SM'91--FÕ99) received his B.S. (with highest honors), M.S., and Ph.D. degrees in Electrical and Computer Engineering from University of Illinois at Urbana-Champaign in 1981, 1983, and 1986, respectively. From 1981 to 1985, he was a Research Assistant in the Coordinated Science Laboratory, University of Illinois.  In August 1985, he joined the faculty of Rice University, Houston, Texas, where he is now the J.S. Abercrombie Professor, and Chair of the Department of Electrical and Computer Engineering.
In addition, he holds an Academy of Finland Distinguished Visiting Professorship appointment (FiDiPro) at the Center for Wireless Communication (CWC) in the University of Oulu, Oulu, Finland. He has served as the founding director of Rice's Center for Multimedia Communications from 1998 till 2006.
His research interests are in the areas of communication theory, information theory, signal processing, and their applications to wireless communication, wireless networks, and neuro-engineering with emphasis on closed-loop real-time neuromodulation and brain stimulation.

Dr. Aazhang is a Fellow of IEEE and AAAS, a distinguished lecturer of IEEE Communication Society, and also a recipient of 2004 IEEE Communication Society's Stephen O. Rice best paper award for a paper with A. Sendonaris and E. Erkip. He has been listed in the Thomson-ISI Highly Cited Researchers and has been keynote and plenary speaker of several conferences. He has served as the co-chair of the Technical Program Committee of 2001 Multi-Dimensional and Mobile Communication (MDMC) Conference in Pori, Finland, the chair of the Technical Program Committee for 2005 Asilomar Conference, Monterey, CA, the co-chair of the Technical Program Committee of International Workshop on Convergent Technologies (IWCT), Oulu, Finland in 2005, guest editor for IEEE Journal on Selected Areas of Communication special issue on relay and cooperative communication in 2006, the general chair of the 2006 Communication Theory Workshop, Dorado, Puerto Rico, and the co-general chair of 2010 International Symposium on Information Theory (ISIT), in Austin, Texas.
\end{IEEEbiography}

\end{document}